\documentclass[12pt]{article}
\usepackage{caption}
\usepackage{amssymb}
\usepackage{amsmath}
\usepackage{hyperref}
\usepackage{graphicx}
\usepackage{float}
\usepackage{bm}
\usepackage{latexsym}
\usepackage{subfig}
\usepackage{mathtools}
\usepackage{booktabs}
\usepackage{tabularx}   
\usepackage{geometry} 
\begin{document}
\title{\bf Thermodynamic Behavior of a 4D Nonminimal Maxwell-AdS Black Hole}
\author{Mehdi Sadeghi\thanks{Corresponding author: Email:  mehdi.sadeghi@abru.ac.ir}\,\,\,  and  \,\,Faramaz Rahmani\thanks{Email:  faramarz.rahmani@abru.ac.ir}\hspace{2mm}\\
	{\small {\em Department of Physics, Faculty of Basic Sciences,}}\\
	{\small {\em Ayatollah Boroujerdi University, Boroujerd, Iran}}
}
\date{\today}
\maketitle

\abstract{In this paper, we derive a black hole solution within the Einstein–Maxwell framework incorporating a non-minimal coupling between the Ricci tensor and the Maxwell field strength tensor, using a perturbative approach. We subsequently explore the thermodynamic phase transitions of the black hole in an extended phase space, analyzing both canonical and grand canonical ensembles. Our findings reveal that the system exhibits Van der Waals-like behavior in both ensembles. Moreover, for sufficiently small values of electric charge and Maxwell potential, the thermodynamics is dominated by a Hawking–Page phase transition.
}\\

\noindent PACS numbers: 11.10.Jj, 11.10.Wx, 11.15.Pg, 11.25.Tq, 04.70.Dy, 04.50.Kd, 04.20.-q \\

\noindent \textbf{Keywords:} Black hole phase transition, AdS black holes, Extended phase space, AdS/CFT duality
\section{Introduction} \label{intro}

Non-minimal gravitational theories propose that the gravitational field is coupled to other fields beyond just the curvature of spacetime. This contrasts with Einstein's theory of general relativity, where the gravitational field is solely determined by the curvature of spacetime. Non-minimal theories have been explored as alternative frameworks for understanding gravitational phenomena at different scales. These theories investigate a variety of intriguing problems that go beyond the scope of traditional general relativity. By modifying Einstein’s equations, they aim to address issues such as dark energy and dark matter, which remain unexplained within the standard cosmological model. They also explore singularities and high-curvature behavior, potentially resolving issues in black hole and Big Bang scenarios~\cite{huang2016,balakin2008}.

Additionally, non-minimal theories examine the coupling of gravity with other fundamental fields, such as scalar fields, electromagnetic fields, and gauge fields, to better understand phenomena like cosmic inflation, anisotropies in the cosmic microwave background, and gravitational waves~\cite{azevedo2018}. By introducing additional degrees of freedom and interactions, these theories aim to provide a more comprehensive framework for understanding the universe's fundamental forces and the evolution of its structure at both cosmological and quantum scales.

The applications of non-minimal gravitational theories are vast and varied. In cosmology, they offer potential explanations for the accelerated expansion of the universe attributed to dark energy and provide alternative models for the inflationary period of the early universe. In astrophysics, these theories assist in modeling the behavior of compact objects like neutron stars and black holes, offering insights into their stability, structure, and radiation mechanisms. Non-minimal couplings also play a significant role in understanding the propagation of gravitational waves and their interactions with matter, potentially leading to new detection methods and improved sensitivity in gravitational wave observatories. Furthermore, these theories contribute to high-energy physics by exploring the unification of fundamental forces and the influence of gravity on quantum fields. Through these applications, non-minimal gravitational theories not only address unresolved questions in modern physics but also open new avenues for experimental and observational investigation.

Non-minimal theories can be classified into several key categories:
\begin{enumerate}
    \item The coupling of scalar fields, such as in the Brans–Dicke theory, where the scalar field is coupled to the Ricci scalar~\cite{Brans:1962zz}.
    \item The non-minimal coupling of the electromagnetic field to curvature, known as non-minimal Einstein–Maxwell models~\cite{Balakin:2005fu, Bamba:2008ja, Horndeski:1978ca, Mueller-Hoissen:1988cpx, Balakin:2010ar}.
    \item The coupling between Yang–Mills gauge fields and the gravitational field, forming Einstein–Yang–Mills models with $SU(n)$ symmetry~\cite{hashimoto1999, bjoraker2000, Sadeghi:2023pdx, Rahmani:2024cfi, Sadeghi:2023hxd, Sadeghi:2022bsh}.
    \item The extension to include the Higgs field, resulting in Einstein–Yang–Mills–Higgs models~\cite{Balakin:2007nw, alvarez2019}.
    \item The coupling between the axion field and the Einstein–Maxwell system, forming Einstein–Maxwell–axion models~\cite{Balakin:2009rg, banerjee2018}.
\end{enumerate}

In the realm of modern physics, the study of the thermodynamics of Anti-de Sitter (AdS) black holes has emerged as a pivotal area of research, unveiling profound insights into the nature of gravity and quantum mechanics. These investigations show that AdS black holes not only serve as critical objects in understanding spacetime geometry but also play a central role in the AdS/CFT correspondence—a foundational principle in theoretical physics. Through the exploration of phase transitions akin to those observed in classical thermodynamic systems, such as the Hawking–Page transition, these studies bridge the gap between macroscopic and microscopic phenomena. The intricate behavior of AdS black holes, including their stability and response to external perturbations, provides crucial clues to the underlying principles of quantum gravity and the behavior of strongly coupled systems. Consequently, the thermodynamic properties of AdS black holes are not merely of academic interest but are instrumental in advancing our understanding of the fundamental laws governing the universe~\cite{hawking1983, cai2002, hartnoll2008, gibbons2005, larsen1996}.

In the thermodynamic study of black holes in AdS space, the concept of extended phase space revolutionizes our understanding by treating the cosmological constant as a dynamic thermodynamic variable, analogous to pressure in classical thermodynamics~\cite{bairagya2021, hendi2018, rehan2019, dolan2011, Robinson2015, Robinson2024, abdusattar2023}. This approach enables the definition of new thermodynamic quantities such as volume, allowing for a more intricate and comprehensive analysis of black hole thermodynamics. The extended phase space framework facilitates the study of critical phenomena, phase transitions, and stability conditions in black holes, providing deeper insights into their behavior and into the AdS/CFT correspondence—thus bridging macroscopic thermodynamic properties with quantum gravity theories~\cite{Balakin:2015gpq, witten1998, wei2012, caldarelli2000, kubiznak2012, Maldacena:1997re, Aharony:1999ti, Zeng:2015wtt, Zeng:2015tfj, He:2016fiz}.

AdS black holes exhibit several types of thermodynamic phase transitions, each providing unique insights into their properties and behavior. One prominent example is the Hawking–Page phase transition, which describes the transition between thermal AdS space and a large AdS black hole, analogous to the phase transition between gas and liquid in classical thermodynamics. Another significant example is the Van der Waals–like phase transition observed in charged AdS black holes, where the system undergoes a transition between small and large black hole phases, similar to the liquid–gas transition in Van der Waals fluids.

The investigation of thermodynamic phase transitions in black holes containing the Maxwell term is of significant importance in theoretical physics. These studies provide deeper insights into the intricate relationship between gravity and gauge fields, enhancing our understanding of unified field theories. By examining how gauge fields, represented by the Maxwell term, influence the thermodynamic properties and stability of black holes, researchers can explore critical phenomena that mirror those in standard thermodynamic systems. This analysis not only improves our comprehension of the microstates underlying black hole entropy but also sheds light on the AdS/CFT correspondence, where gauge field dynamics play a central role. Moreover, understanding these phase transitions aids in clarifying the complex interactions between different fundamental forces, potentially guiding the development of more complete models of quantum gravity. 

The coupling of gravity and matter fields—particularly the coupling between the Ricci tensor and the Maxwell tensor—is of paramount importance in black hole thermodynamics. This interaction enriches our understanding of the complex dynamics within black hole systems, where gravitational and gauge field effects intertwine. By incorporating the Ricci tensor, which encapsulates spacetime curvature, with the Maxwell tensor, representing the field strength of gauge fields, researchers can investigate how these combined forces affect black hole behavior. This coupling leads to more comprehensive models that capture the subtleties of black hole thermodynamics, including phase transitions and stability. Analyzing these interactions provides critical insights into the microstates contributing to black hole entropy and the fundamental mechanisms of quantum gravity.

We study the thermodynamic phase transition of the black hole solution in both canonical and grand canonical ensembles. While the canonical ensemble provides valuable insights into energy exchange and thermal behavior, the grand canonical ensemble extends this by incorporating charge and particle number variability. This leads to a more comprehensive and nuanced picture of black hole thermodynamics. The grand canonical ensemble allows for fluctuations in the number of particles (or charge, in the case of black holes), offering insights into how black holes respond to environmental changes such as varying charge or matter influx. It is particularly useful in the context of the AdS/CFT correspondence, where it helps model the dual conformal field theory with variable particle numbers or charges, enriching the theoretical framework.

In Section~\ref{sec2}, the black hole solution will be derived through a perturbative approach. Then, in Sections~\ref{sec3} and~\ref{sec4}, the thermodynamics of the solution will be investigated.

\section{ Perturbative Solution of a Nonminimal Maxwell-AdS Black Hole}
\label{sec2}

In the introduction of this paper, we highlighted the significance of non-minimal models in the thermodynamic analysis of black holes. Our non-minimal model is defined by the following bulk action:
\begin{align}\label{action}
S &= \int d^{4} x \sqrt{-g} \bigg[ \frac{1}{2}(R - 2\Lambda) - \tfrac{1}{4} F_{\mu \nu} F^{\mu \nu} + \nonumber \\
& \quad \frac{\lambda}{2} F_{\alpha \beta} F_{\mu \nu} \left( g^{\nu \beta} R^{\mu \alpha} - g^{\nu \alpha} R^{\mu \beta} - g^{\mu \beta} R^{\nu \alpha} + g^{\mu \alpha} R^{\nu \beta} \right) \bigg],
\end{align}
where \( R \) is the Ricci scalar, \( \Lambda = -\frac{3}{l^2} \) denotes the cosmological constant, and \( l \) is the AdS radius. The Maxwell invariant is defined as \( \mathcal{F} = F_{\mu \alpha} F^{\mu \alpha} \), where \( F_{\mu \nu} \) is the field strength tensor associated with the gauge field \( A_\nu \). The field strength tensor takes the form
\begin{align} \label{YM}
F_{\mu \nu} = \partial_\mu A_\nu - \partial_\nu A_\mu,
\end{align}
which encodes the antisymmetric derivatives of the gauge potential. The Ricci tensor is denoted by \( R^{\mu \alpha} \). In our model, the parameter \( \lambda \), which has the dimension \([\lambda] = [L]^2\), is generally arbitrary and serves as a phenomenological constant characterizing the non-minimal coupling between the Maxwell field and the Ricci tensor \cite{Balakin:2005fu}. This coupling modifies the standard Maxwell theory and enables a richer thermodynamic and phase structure for black holes in the AdS background. The coupling term is inspired by Ref.~\cite{Balakin:2015gpq}, including all possible permutations of the metric $g_{\mu \nu}$  to ensure completeness.

Our aim is to obtain black hole solutions within this non-minimal model. For this purpose, we adopt a general ansatz for the spacetime metric in the following form:
\begin{equation}\label{metric}
ds^{2} = -f(r)e^{-2H(r)} dt^{2} + \frac{dr^{2}}{f(r)} + r^2 d\theta^2 + r^2 \sin^2 \theta\, d\phi^2,
\end{equation}
where the factor \( e^{-2H(r)} \) reflects the impact of the non-minimal Maxwell coupling.

The gauge field is assumed to have the form:
\begin{equation}  
A_{\mu} = \left(h(r), 0, 0, 0\right),  
\end{equation}  
implying that the only non-zero component is \( A_0 \). Consequently, the field strength tensor \( F_{\mu \nu} \) becomes:  
\begin{equation}   
\label{YM2}  
F_{\mu \nu} = \partial_{\mu} A_{\nu} - \partial_{\nu} A_{\mu} =  \left(  
\begin{array}{cccc}  
0 & -h'(r) & 0 & 0 \\
h'(r) & 0 & 0 & 0 \\
0 & 0 & 0 & 0 \\
0 & 0 & 0 & 0 \\
\end{array}  
\right).   
\end{equation}  

Furthermore, we have:  
\begin{equation}  
\mathcal{F} = F_{\mu \nu} F^{\mu \nu} = -2 e^{2 H(r)} h'(r)^2.  
\end{equation}

Varying the action \eqref{action} with respect to the metric \( g_{\mu \nu} \) yields the field equation:
\begin{equation}\label{EOM1}
R_{\mu \nu } - \tfrac{1}{2} g_{\mu \nu } R + \Lambda g_{\mu \nu } =  T^{\text{(eff)}}_{\mu \nu },
\end{equation}
where
\begin{equation}
T^{\text{(eff)}}_{\mu \nu } = T^{\text{(M)}}_{\mu \nu } + \lambda T^{(I)}_{\mu \nu },
\end{equation}
with
\begin{equation}
T^{\text{(M)}}_{\mu \nu } =  F_{\mu }{}^{\alpha } F_{\nu \alpha } - \tfrac{1}{4} g_{\mu \nu } F_{\alpha \beta } F^{\alpha \beta },
\end{equation}
\begin{align}\label{tt}
-2 T^{(I)}_{\mu \nu } &= 2 F_{\mu }{}^{\alpha } F_{\nu }{}^{\beta } R_{\alpha \beta } -  F_{\alpha }{}^{\gamma } F^{\alpha \beta } g_{\mu \nu } R_{\beta \gamma } - 2 F_{\alpha }{}^{\beta } F_{\nu }{}^{\alpha } R_{\mu \beta } \nonumber \\ 
&\quad - 2 F_{\alpha }{}^{\beta } F_{\mu }{}^{\alpha } R_{\nu \beta } + F_{\nu }{}^{\alpha } \nabla_{\beta }\nabla^{\beta }F_{\mu \alpha } + F_{\mu }{}^{\alpha } \nabla_{\beta }\nabla^{\beta }F_{\nu \alpha } \nonumber \\ 
&\quad + F^{\alpha \beta } g_{\mu \nu } \nabla_{\beta }\nabla_{\gamma }F_{\alpha }{}^{\gamma } + F_{\nu }{}^{\alpha } \nabla_{\beta }\nabla_{\mu }F_{\alpha }{}^{\beta } + F^{\alpha \beta } \nabla_{\beta }\nabla_{\mu }F_{\nu \alpha } \nonumber \\ 
&\quad + F_{\mu }{}^{\alpha } \nabla_{\beta }\nabla_{\nu }F_{\alpha }{}^{\beta } + F^{\alpha \beta } \nabla_{\beta }\nabla_{\nu }F_{\mu \alpha } + 2 \nabla_{\beta }F_{\mu }{}^{\alpha } \nabla^{\beta }F_{\nu \alpha } \nonumber \\ 
&\quad + g_{\mu \nu } \nabla_{\beta }F^{\alpha \gamma } \nabla_{\gamma }F_{\alpha }{}^{\beta } -  g_{\mu \nu } \nabla_{\alpha }F^{\alpha \beta } \nabla_{\gamma }F_{\beta }{}^{\gamma } \nonumber \\ 
&\quad + F^{\alpha \beta } g_{\mu \nu } \nabla_{\gamma }\nabla_{\beta}F_{\alpha }{}^{\gamma } + \nabla_{\beta }F_{\nu }{}^{\alpha } \nabla_{\mu }F_{\alpha }{}^{\beta } + \nabla_{\beta }F^{\alpha \beta } \nabla_{\mu }F_{\nu \alpha } \nonumber \\ 
&\quad + \nabla_{\beta }F_{\mu }{}^{\alpha } \nabla_{\nu }F_{\alpha }{}^{\beta } + \nabla_{\beta }F^{\alpha \beta } \nabla_{\nu }F_{\mu \alpha }.
\end{align}

The effective energy-momentum tensor \( T^{\text{(eff)}}_{\mu \nu} \), which contains contributions from both the standard Maxwell term \( T^{\text{(M)}}_{\mu \nu} \) and the interaction term \( T^{(I)}_{\mu \nu} \), appears on the right-hand side of Einstein’s equations. 

The modified Maxwell equations are obtained by varying the action \eqref{action} with respect to the gauge field \( A_{\mu} \):
\begin{equation}\label{EOM-YM}
\nabla_{\nu}\left( -\tfrac{1}{2}  F^{\mu \nu } -2 \lambda F^{\nu \alpha } R^{\mu }{}_{\alpha } +2 \lambda F^{\mu \alpha } R^{\nu }{}_{\alpha }\right) = 0.
\end{equation}

The \( tt \)-component of Einstein's field equations is obtained from Eq.~(\ref{EOM1}) and is as follows:
\begin{equation}\label{t1}
2 - 2 r f'(r) - 2 f(r) - r^2 e^{2 H(r)} h'(r)^2 - 2 \Lambda r^2 + \lambda B_1(r) = 0,
\end{equation}
where
\begin{align}
B_1(r) =\ &32 r^2 e^{2 H(r)} f'(r) h'(r)^2 H'(r) - 8 r^2 e^{2 H(r)} f''(r) h'(r)^2 + 8 r^2 e^{2 H(r)} f'(r) h'(r) h''(r) \nonumber\\
&- 8 r e^{2 H(r)} f'(r) h'(r)^2 + 64 r^2 f(r) e^{2 H(r)} h'(r) h''(r) H'(r) + 32 r^2 f(r) e^{2 H(r)} h'(r)^2 H''(r) \nonumber\\
&+ 16 r^2 f(r) e^{2 H(r)} h'(r)^2 H'(r)^2 + 64 r f(r) e^{2 H(r)} h'(r)^2 H'(r) + 48 r f(r) e^{2 H(r)} h'(r) h''(r) \nonumber\\
&+ 16 r^2 f(r) h^{(3)}(r) e^{2 H(r)} h'(r) + 8 f(r) e^{2 H(r)} h'(r)^2 + 16 r^2 f(r) e^{2 H(r)} h''(r)^2.
\end{align}

For \( \lambda = 0 \), Eq.~(\ref{t1}) reduces to:
\begin{equation}\label{ttq2}
2 - 2 r f'(r) - 2 f(r) - r^2 e^{2 H(r)} h'(r)^2 - 2 \Lambda r^2 = 0.
\end{equation}

The \( rr \)-component of Einstein's equations is given by:
\begin{equation}\label{r1}
2 r f'(r) + 2 f(r) \left(1 - 2 r H'(r)\right) + r^2 e^{2 H(r)} h'(r)^2 + 2 \Lambda r^2 - 2 + \lambda B_2(r) = 0,
\end{equation}
where
\begin{align}
B_2(r) =\ &-32 r^2 e^{2 H(r)} f'(r) h'(r)^2 H'(r) + 8 r^2 e^{2 H(r)} f''(r) h'(r)^2 - 8 r^2 e^{2 H(r)} f'(r) h'(r) h''(r) \nonumber\\
&+ 8 r e^{2 H(r)} f'(r) h'(r)^2 + 16 r^2 f(r) e^{2 H(r)} h'(r) h''(r) H'(r) - 16 r^2 f(r) e^{2 H(r)} h'(r)^2 H''(r) \nonumber\\
&+ 32 r^2 f(r) e^{2 H(r)} h'(r)^2 H'(r)^2 - 16 r f(r) e^{2 H(r)} h'(r)^2 H'(r) \nonumber\\
&- 16 r f(r) e^{2 H(r)} h'(r) h''(r) - 8 f(r) e^{2 H(r)} h'(r)^2.
\end{align}

For \( \lambda = 0 \), Eq.~(\ref{r1}) simplifies to:
\begin{equation}\label{rrq2}
2 r f'(r) + 2 f(r) \left(1 - 2 r H'(r)\right) + r^2 e^{2 H(r)} h'(r)^2 + 2 \Lambda r^2 - 2 = 0.
\end{equation}

By subtracting Eq.~(\ref{rrq2}) from Eq.~(\ref{ttq2}), we obtain:
\begin{equation}
4 r f(r) H'(r) = 0.
\end{equation}
This implies that \( H(r) \) must be a constant when \( \lambda = 0 \). So, we set $H_0=0$ for simplicity.

The \( \theta\theta \)-component of Eq.~(\ref{EOM1}) reads:
\begin{align}\label{xx}
&2 f'(r) - 3 r f'(r) H'(r) + r f''(r) + 2 r f(r) H'(r)^2 - 2 f(r) H'(r) - 2 r f(r) H''(r) \nonumber\\
&- r e^{2 H(r)} h'(r)^2 + \lambda B_3(r) = 0,
\end{align}
where,
\begin{align}
B_3(r) =\ &4 r e^{2 H(r)} f'(r) h'(r)^2 H'(r) - 4 r e^{2 H(r)} f''(r) h'(r)^2 - 8 r e^{2 H(r)} f'(r) h'(r) h''(r) \nonumber\\
&- 8 e^{2 H(r)} f'(r) h'(r)^2 - 24 r f(r) e^{2 H(r)} h'(r) h''(r) H'(r) - 16 r f(r) e^{2 H(r)} h'(r)^2 H'(r)^2 \nonumber\\
&- 8 f(r) e^{2 H(r)} h'(r)^2 H'(r) - 16 f(r) e^{2 H(r)} h'(r) h''(r) \nonumber\\
&- 8 r f(r) h^{(3)}(r) e^{2 H(r)} h'(r) - 8 f(r) e^{2 H(r)} h''(r)^2.
\end{align}

For \( \lambda = 0 \), Eq.~(\ref{xx}) reduces to:
\begin{align}
&2 f'(r) - 3 r f'(r) H'(r) + r f''(r) + 2 r f(r) H'(r)^2 - 2 f(r) H'(r) - 2 r f(r) H''(r) \nonumber\\
&- r e^{2 H(r)} h'(r)^2 = 0.
\end{align}

The \( \phi\phi \)-component of the gravitational field equations is identical to Eq.~(\ref{xx}).

We begin by solving the Maxwell equations (\ref{EOM-YM}) to obtain the function \( h(r) \):
\begin{equation}\label{Maxwell}
h(r) = \int^{r} \frac{e^{-H(u)}}{\mathcal{B}(u)} \, du,
\end{equation}
where
\begin{equation}\label{B}
\mathcal{B}(u) = \lambda u \left[4 f'(u)\left(3 u H'(u) - 2\right) - 4 u f''(u) + 8 f(u)\left(-u H'(u)^2 + H'(u) + u H''(u)\right)\right] - u^2.
\end{equation}

Since an exact solution to the field equations is not tractable, we proceed by solving the system perturbatively to first order in the coupling parameter \( \lambda \). Accordingly, we expand the functions \( f(r) \), \( H(r) \), and \( h(r) \) as:
\begin{align}
f(r) &= f_0(r) + \lambda f_1(r), \label{f} \\
H(r) &= H_0(r) + \lambda H_1(r), \label{H} \\
h(r) &= h_0(r) + \lambda h_1(r). \label{h}
\end{align}
We can determine \( f_0(r) \) and \( h_0(r) \) by utilizing the Einstein equation components in the case of zero non-minimal coupling, i.e. $\lambda=0$.\\ In this regard we have:
\begin{equation}
f_0(r) = 1 - \frac{2 m_0}{r} + \frac{r^2}{l^2} + \frac{Q^2}{2 r^2},
\end{equation}
where the constant \( m_0 \) is chosen such that \( f_0(r) \) vanishes at the horizon radius \( r = r_h \):
\begin{equation}\label{m}
m_0 = \frac{r_h}{2} + \frac{r_h^3}{2 l^2} + \frac{Q^2}{4 r_h}.
\end{equation}
Substituting \( m_0 \) into the expression for \( f_0(r) \) yields the form:
\begin{equation}
f_0(r) = \left(1 - \frac{r_h}{r}\right) + \frac{r^2}{l^2} \left(1 - \frac{r_h^3}{r^3}\right) + \frac{Q^2}{2 r} \left(\frac{1}{r} - \frac{1}{r_h}\right).
\end{equation}
This choice ensures that \( f_0(r_h) = 0 \) at zeroth order in \( \lambda \). A similar condition will be imposed on \( h_0(r) \) at the horizon. It is worth noting that the horizon radius will generally be perturbed at higher orders in \( \lambda \), and we shall analyze this effect in due course.

Using relation (\ref{Maxwell}) and (\ref{B}) with $\lambda=0$, we obtain
\begin{equation}\label{h0}
h_0(r) = Q \left(\frac{1}{r} - \frac{1}{r_h}\right),
\end{equation}
where $h_0(r)$ vanishes on the horizon at the zeroth order in $\lambda$.

By substituting Eqs.~(\ref{f}), (\ref{H}), and (\ref{h}) into Eqs.~(\ref{t1}) and (\ref{r1}), and subtracting them up to first order in \(\lambda\), we obtain
\begin{equation}
4 r f_0(r) \left(4 h_0'(r) \left(2 h_0''(r) + r h_0^{(3)}(r)\right) + 4 r h_0''(r)^2 - H_1'(r)\right) = 0.
\end{equation}
Solving this equation for \(H_1(r)\), we find
\begin{equation}\label{H1}
H_1(r) = 4 \int^r \left(h_0'(u) \left(u h_0^{(3)}(u) + 2 h_0''(u)\right) + u h_0''(u)^2 \right) du.
\end{equation}
By substituting \(h_0(r)\) into Eq.~(\ref{H1}), we obtain
\begin{equation}
H_1(r) = -\frac{6 Q^2}{r^4},
\end{equation}
where the constant of integration has been set to zero without loss of generality.

By inserting Eqs.(\ref{f}, \ref{H}, \ref{h}) into Eq.(\ref{Maxwell}), and considering the first-order term in $\lambda$, we obtain $h_1(r)$ as follows:\\
\begin{equation}\label{h1}
h_1(r) = Q\int^r \frac{8 f_0'(u) + 4 u f_0''(u) + u H_1(u)}{ u^3} du + C_1.
\end{equation}
Evaluating the integral and using \( l^2 = -\frac{3}{\Lambda} \), we find:
\begin{equation}
h_1(r) = \frac{2 Q^3}{5 r^5} + \frac{8 Q \Lambda}{r} + C_1.
\end{equation}

Now, by using either the \(rr\)- or \(tt\)-component of Einstein's field equations up to first order in \(\lambda\), we can derive a differential equation that yields the function \(f_1(r)\) in the form of the following integral:
\begin{equation}\label{f1}
f_1(r) = \frac{1}{r} \left( \int^r E(u) \, du + C_2 \right),
\end{equation}
where
\begin{align}
E(u) ={}& -4 u^2 f_0''(u) h_0'(u)^2 + 4 u^2 f_0'(u) h_0'(u) h_0''(u) - 4 u f_0'(u) h_0'(u)^2 \nonumber \\
&+ 8 u^2 f_0(u) h_0^{(3)}(u) h_0'(u) + 8 u^2 f_0(u) h_0''(u)^2 + 24 u f_0(u) h_0'(u) h_0''(u) \nonumber \\
&+ 4 f_0(u) h_0'(u)^2 - \frac{1}{2} u^2 h_0'(u) h_1'(u) - u^2 H_1(u) h_0'(u)^2.
\end{align}

After performing the integration, we obtain
\begin{equation}
f_1(r) = -\frac{12 Q^2}{r^4} - \frac{23 Q^4}{5 r^6} + \frac{20 Q^2 m_0}{r^5} + \frac{16 Q^2 \Lambda}{3 r^2} + \frac{C_2}{r}.
\end{equation}

We now invoke the concept of a perturbed black hole horizon to determine the integration constants \( C_1 \) and \( C_2 \), which appear in the gauge and metric functions \( h(r) \) and \( f(r) \), respectively. This technique not only ensures the physical consistency of our perturbative solution but also sheds light on the subtle ways in which the horizon structure responds to non-minimal couplings.

To begin, we examine the gauge function \( h(r) \), given by
\begin{equation}\label{hnew}
h(r) = Q\left(\frac{1}{r} - \frac{1}{R}\right) + \left(\frac{8 Q \Lambda}{r} + \frac{2 Q^{3}}{5 r^{5}} + C_1 \right)\lambda,
\end{equation}
where \( C_1 \) is yet to be determined. The horizon of the black hole is defined by the vanishing of this function. When non-minimal coupling is present, the horizon position may shift. To capture this effect perturbatively, we define a shifted horizon radius as
\begin{equation}
r_h^{\prime} = r_h + \lambda r_h^{(1)},
\end{equation}
where \( r_h \) is the unperturbed horizon, and \( \lambda r_h^{(1)} \) is the first-order correction.

Applying the horizon condition \( h(r_h^{\prime}) = 0 \) and expanding in \( \lambda \), we obtain
\begin{align*}
h(r_h + \lambda r_h^{(1)}) &= h_0(r_h) + \lambda \left[ h_0^{\prime}(r_h) r_h^{(1)} + h_1(r_h) \right] + \mathcal{O}(\lambda^2) = 0,
\end{align*}
from which the correction \( r_h^{(1)} \) is found:
\begin{equation}
r_h^{(1)} = -\frac{h_1(r_h)}{h_0^{\prime}(r_h)}.
\end{equation}

This leads to an explicit expression for the perturbed horizon:
\begin{equation}\label{ph}
r_h^{\prime} = r_h - \lambda \left( \frac{h_1(r_h)}{h_0^{\prime}(r_h)} \right).
\end{equation}

Imposing the condition \( h(r_h^{\prime}) = 0 \) and expanding consistently in \( \lambda \) allows us to solve for the integration constant:
\begin{equation*}
C_1 = -\frac{8 Q \Lambda}{r_h} - \frac{2 Q^{3}}{5 r_h^{5}}.
\end{equation*}

Substituting back into Eq.~\eqref{hnew}, we obtain a fully determined form for the gauge function:
\begin{equation}\label{hnew2}
h(r) = \left(\frac{1}{r} - \frac{1}{r_h}\right) Q + \left[\left(\frac{8 \Lambda}{r} - \frac{8 \Lambda}{r_h}\right) Q + \left(\frac{2}{5 r^{5}} - \frac{2}{5 r_h^{5}}\right) Q^{3}\right] \lambda.
\end{equation}

An intriguing consequence is that this function still vanishes at the original, unperturbed horizon \( r = r_h \), implying that, to first order in \( \lambda \), the location of the black hole horizon remains unchanged. This behavior, however, is model-dependent and not guaranteed in general; higher-order corrections or alternate couplings may induce a genuine horizon shift, a topic that merits further exploration.

To ensure internal consistency of our perturbative scheme, the horizon correction derived from \( h(r) \) must agree with that obtained from the metric function \( f(r) \). This yields a matching condition between the first-order corrections:
\begin{equation}\label{con}
\frac{h_1(r_h)}{h_0^{\prime}(r_h)} = \frac{f_1(r_h)}{f_0^{\prime}(r_h)}.
\end{equation}

Rather than directly imposing \( f(r_h) = 0 \), we find that this condition provides a more elegant and reliable route to determine the remaining integration constant \( C_2 \) in \( f(r) \). Upon solving and substituting the result, the final form of the metric function up to \( \mathcal{O}(\lambda) \) reads:
\begin{equation}
\begin{aligned}
f(r) =\; & 1 - \frac{\Lambda r^2}{3} 
+ \left( \frac{r_h^3 \Lambda}{3} - \frac{Q^2}{2r_h} - r_h \right) \frac{1}{r} 
+ \frac{Q^2}{2 r^2} \\[5pt]
& +  \left[ 
\left( -\frac{2 Q^2 \Lambda}{r_h} 
- \frac{2 Q^4}{5 r_h^5} 
+ \frac{2 Q^2}{r_h^3} 
\right) \frac{1}{r} 
+ \frac{16 Q^2 \Lambda}{3 r^2} 
- \frac{12 Q^2}{r^4} \right. \\[5pt]
& \left. \quad 
+ \left( 
- \frac{10 Q^2 r_h^3 \Lambda}{3} 
+ \frac{5 Q^4}{r_h} 
+ 10 Q^2 r_h 
\right) \frac{1}{r^5} 
- \frac{23 Q^4}{5 r^6} 
\right]\lambda.
\end{aligned}
\end{equation}

This expression fully encapsulates the backreaction of the non-minimally coupled Maxwell field on the black hole geometry at linear order in the coupling constant.

To assess the regularity of the black hole solution, we examine the Kretschmann scalar, defined as
\begin{equation}\label{KS1}
K = R_{\mu \nu \rho \sigma} R^{\mu \nu \rho \sigma},
\end{equation}
which encapsulates the full contraction of the Riemann tensor and provides a coordinate-independent measure of spacetime curvature.

For the class of metrics under consideration, the Kretschmann scalar evaluates to
\begin{equation}\label{KS2}
K = \frac{4(f - 1)^2}{r^4} +
\frac{1}{r^2} \left[2 (f')^2 + 2 \left(f' - 2 f H'\right)^2 \right] +
\left(f'' + 2 f (H' - H'') - 3 f' H' \right)^2.
\end{equation}

This scalar serves as a powerful diagnostic for identifying geometric pathologies, particularly curvature singularities. Upon explicit computation, it becomes evident that the only divergence in \( K \) occurs at \( r = 0 \), indicating a physical singularity at the origin—an expected feature in many spherically symmetric black hole solutions.

In the asymptotic regime \( r \to \infty \), the scalar approaches the constant value
\begin{equation}
K \to \frac{8 \Lambda^2}{3},
\end{equation}
which is characteristic of a spacetime with negative constant curvature. This limiting behavior confirms that the geometry asymptotically approaches that of anti-de Sitter (AdS) space. The finiteness of the Kretschmann scalar at large \( r \) ensures the absence of curvature singularities in the asymptotic region, reinforcing the physical viability of the solution.

\begin{figure}[H]
\centering
\includegraphics[width=8cm]{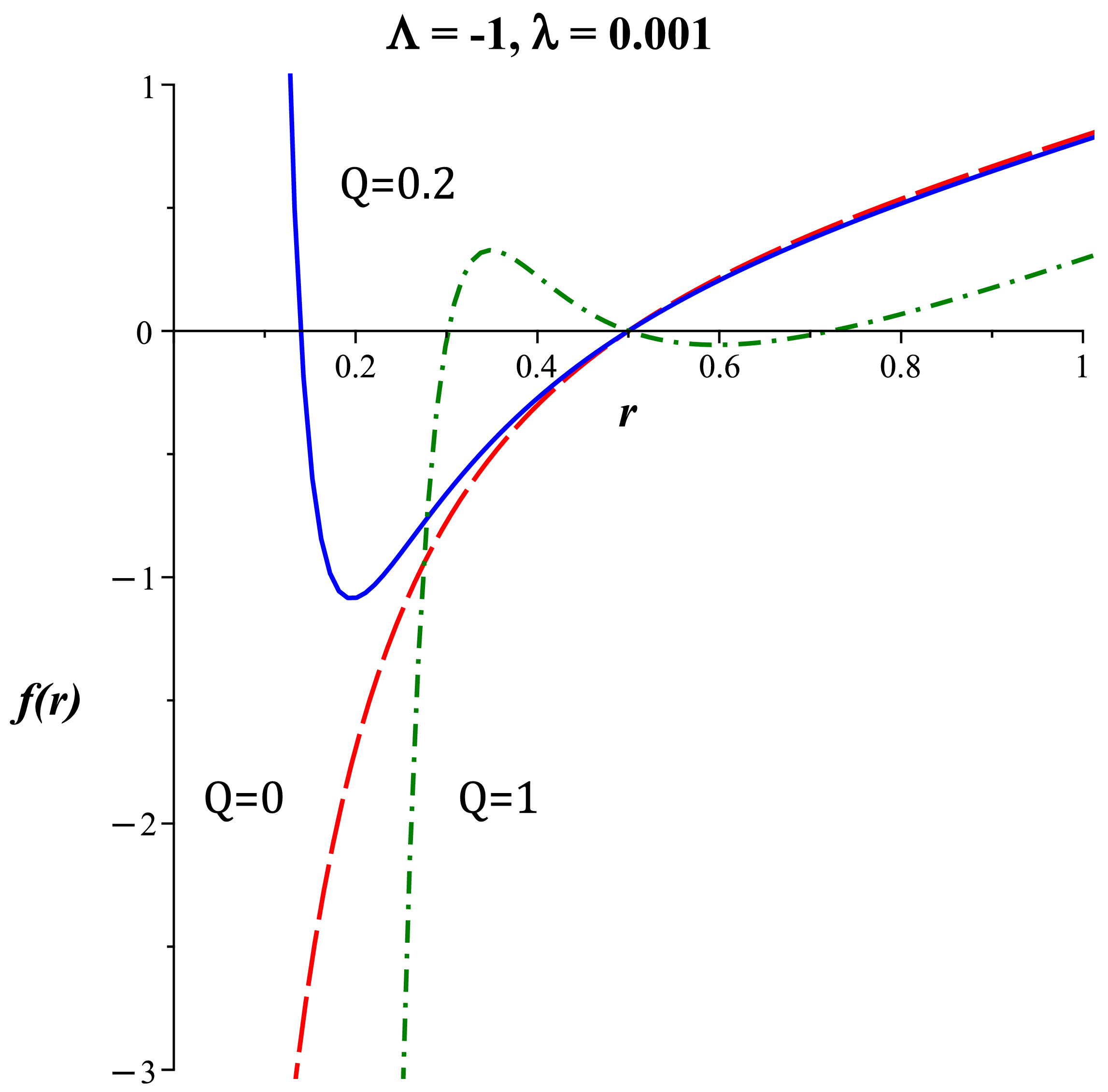}
\caption{Profile of the metric function \( f(r) \) for  different  Maxwell charges \( Q \), showing the emergence of multiple horizons. As the charge increases, the number of roots of \( f(r) \) also increases. The analysis reveals that up to three distinct roots may emerge for sufficiently large values of \( Q \).}
\label{fig:1}
\end{figure}

The behavior of the metric function \( f(r) \) for various values of the Maxwell charge \( Q \) is illustrated in Fig.~\ref{fig:1}. This figure captures how the presence and magnitude of the electromagnetic field affect the geometry of the black hole, offering insights into the resulting horizon structure.

In the absence of electromagnetic charge (\( Q = 0 \)), the function \( f(r) \) admits a single real root, corresponding to a unique event horizon. This behavior is indicative of a non-extremal black hole with a relatively simple causal structure. The lack of additional horizons in this case reflects the minimal influence of matter fields on the spacetime geometry.

As the Maxwell charge increases, the profile of \( f(r) \) evolves significantly. Beyond a critical charge threshold, additional roots appear—signifying the formation of inner and outer horizons. This feature is reminiscent of charged black holes such as Reissner--Nordström and signals the transition to more intricate spacetime configurations. 

These results underscore the profound influence of the Maxwell field on black hole geometry. Not only does it modify the number and location of horizons, but it also affects key thermodynamic quantities such as temperature, entropy, and specific heat—ultimately shaping the phase structure and stability characteristics of the system.

By expanding the time-time component of the metric, \( g_{tt} = e^{-2H(r)} f(r) \), in a perturbative series in the coupling parameter \( \lambda \), and comparing the resulting expression with the standard asymptotic form of an asymptotically AdS spacetime,
\begin{equation}
g_{tt} = 1 - \frac{\Lambda r^2}{3} - \frac{2m}{r} + \cdots,
\end{equation}
we extract an expression for the ADM mass of the black hole up to first order in \( \lambda \):
\begin{equation}\label{mass}
m = \frac{r_h}{2} - \frac{r_h^3 \Lambda}{6} + \frac{Q^2}{4 r_h} + \left( \frac{Q^2 \Lambda}{r_h} + \frac{Q^4}{5 r_h^5} - \frac{Q^2}{r_h^3} \right) \lambda.
\end{equation}

This result clearly demonstrates the contribution of the non-minimal coupling between the electromagnetic field and curvature to the black hole mass. In the limit \( \lambda \to 0 \), the expression reduces to the familiar ADM mass of a charged AdS black hole (i.e., the AdS–Maxwell solution), serving as a consistency check on our perturbative expansion.

It is important to emphasize that this derivation relies on the assumption of an asymptotically AdS geometry. In spacetimes where this asymptotic structure is modified—either due to higher-order curvature corrections or alternative boundary conditions—the identification of the ADM mass via this expansion may require modification or an alternative approach. Nevertheless, within the present framework, the result in Eq.~\eqref{mass} encapsulates the leading-order correction to the black hole mass due to non-minimal electromagnetic couplings, and provides valuable insight into the interplay between matter fields and gravitational dynamics in extended theories of gravity.
\section{Critical Behavior and Thermodynamic Phase Transition}
\label{sec3}
It is well established that, in Anti-de Sitter (AdS) space without matter content or higher-order gravitational corrections, a black hole solution undergoes a phase transition between thermal AdS space and black hole phases. This phenomenon, known as the \textit{Hawking--Page phase transition}, is a first-order phase transition.

Within this framework, two distinct black hole configurations can arise: a \textit{small, unstable black hole} and a \textit{large, stable black hole}. The phase transition occurs between these states, representing a shift from a regime dominated by thermal AdS space to one governed by a black hole. This behavior reflects the thermodynamic nature of black holes, wherein the smaller black hole possesses higher free energy compared to the larger one, thereby rendering the latter thermodynamically favored and stable~\cite{hawking1983}.

In the following analysis, we explore how the inclusion of matter-related terms in the action~(\ref{action}) modifies the properties of the background spacetime.

To investigate the thermodynamic characteristics of the black hole, we first evaluate the ADM mass, which corresponds to the enthalpy in the extended phase space framework. In four-dimensional models, the ADM mass coincides with the integration constant \( m \) appearing in relation~(\ref{mass}), thus facilitating the computation of enthalpy.

Using the relation \( \Lambda = -8 \pi P \), the enthalpy of the system, accurate to first order in the coupling constant \( \lambda \), is given by:
\begin{equation}\label{ent}
H = \frac{r_h}{2} + \frac{4 \pi P r_h^3}{3} + \frac{Q^2}{4 r_h} + \left(-\frac{8 \pi P Q^2}{r_h} + \frac{Q^4}{5 r_h^5} - \frac{Q^2}{r_h^3} \right) \lambda.
\end{equation}

Now, let us compute the Hawking temperature for this black hole solution using the standard relation~\cite{Hawking:1974rv}:
\begin{equation}\label{Temp}
T = \frac{1}{2 \pi} \left[ \frac{1}{\sqrt{g_{rr}}} \frac{d}{dr} \sqrt{-g_{tt}} \right] \Bigg|_{r = r_h} = \frac{e^{-H(r_h)} f'(r_h)}{4 \pi}.
\end{equation}
Expanding this result to first order in the coupling constant \( \lambda \), we find:
\begin{align}\label{temp1}
T =\ & \frac{1}{4\pi r_h} - \frac{\Lambda r_h}{4\pi} - \frac{Q^2}{8\pi r_h^3} + \left( \frac{Q^2 \Lambda}{2\pi r_h^3} + \frac{Q^2}{2\pi r_h^5} \right) \lambda \notag \\
=\ & \frac{1}{4\pi r_h}+  2 P r_h  - \frac{Q^2}{8\pi r_h^3} + \left( -\frac{4 Q^2 P}{r_h^3} + \frac{Q^2}{2\pi r_h^5} \right) \lambda.
\end{align}

From the temperature expression, the pressure can be obtained by inverting the temperature–horizon radius relation:
\begin{equation}\label{p1}
P = -\frac{8 \pi T r_{h}^{5} + Q^{2} r_{h}^{2} - 2 r_{h}^{4} - 4 Q^{2} \lambda}{16 \pi r_{h}^{2} \left(-r_{h}^{4} + 2 Q^{2} \lambda \right)}.
\end{equation}
Expanding this to first order in \( \lambda \), we obtain:
\begin{equation}\label{p2}
P = \frac{T}{2 r_{h}} + \frac{Q^{2}}{16 \pi r_{h}^{4}} - \frac{1}{8 \pi r_{h}^{2}} + \left( -\frac{Q^{2}}{2 \pi r_{h}^{6}} + \frac{Q^{2} T}{r_{h}^{5}} + \frac{Q^{4}}{8 \pi r_{h}^{8}} \right) \lambda.
\end{equation}

The entropy of the black hole, computed using the thermodynamic identity
\[
S = \int \frac{1}{T} \frac{dH}{dr_h} \, dr_h,
\]
is given to first order in \( \lambda \) by:
\begin{equation}\label{ent1}
S = \pi r_h^2 - \frac{4\pi Q^2 \lambda}{r_h^2}.
\end{equation}
The leading term \( \pi r_h^2 \) corresponds to the Bekenstein–Hawking entropy, while the second term reflects the correction due to the nonminimal matter–curvature coupling.

Importantly, the correction is inversely proportional to the square of the horizon radius, implying that as \( r_h \) decreases, the magnitude of the correction increases. Thus, for sufficiently small black holes, the entropy may increase compared to its classical value. This behavior deviates from classical expectations, where entropy is strictly proportional to area, and suggests that matter-curvature couplings can significantly influence thermodynamic quantities at small scales.

The remaining thermodynamic quantities will be explored in subsequent sections as we analyze the system in both canonical and grand canonical ensembles.
\subsection{Canonical Ensemble Analysis: Thermodynamics and Stability}\label{subsec1}
In the canonical ensemble, the number of particles is held constant. For black hole thermodynamics, this corresponds to fixing the Maxwell charge \( Q \). Consequently, any stability analysis within this ensemble must account explicitly for the dependence of thermodynamic quantities on \( Q \).

The first law of thermodynamics for the system takes the form:
\begin{equation}
    dH(S, P, Q) = T\,dS + V\,dP + \Phi\,dQ,
\end{equation}
where the thermodynamic quantities are defined as follows:
\begin{equation}
\begin{split}
T &= \left( \frac{\partial H}{\partial S} \right)_{P,Q} 
    = \left( \frac{ \frac{\partial H}{\partial r_h} }{ \frac{\partial S}{\partial r_h} } \right)_{P,Q} 
    = 2 r_h P + \frac{1}{4 \pi r_h} - \frac{Q^2}{8 \pi r_h^3} 
    + \left( -\frac{4 Q^2 P}{r_h^3} + \frac{Q^2}{2 \pi r_h^5} \right) \lambda,
\end{split}
\end{equation}
   \begin{equation}
    V = \left( \frac{\partial H}{\partial P} \right)_{S,Q} 
    = \frac{4 \pi r_h^3}{3} - \frac{8 \pi Q^2 \lambda}{r_h},
   \end{equation}
   and
   \begin{equation}\label{fi}
    \Phi = \left( \frac{\partial H}{\partial Q} \right)_{S,P} 
    = \frac{Q}{2 r_h} + \left( -\frac{16 \pi P Q}{r_h} + \frac{4 Q^3}{5 r_h^5} - \frac{2 Q}{r_h^3} \right) \lambda.
   \end{equation}
Here, \( V \) and \( \Phi \) are the conjugate volume and electric potential to the pressure \( P \) and charge \( Q \), respectively.

Based on the approximate solutions within this model, the Smarr relation must be adjusted according to the degree of approximation. To first order in $\lambda$, the modified Smarr relation takes the form:
\begin{equation}
    H = 2TS - 2VP + Q\Phi,
\end{equation}
where $H$ is the enthalpy, $T$ the temperature, $S$ the entropy, $V$ the thermodynamic volume, $P$ the pressure, $Q$ the charge, and $\Phi$ the electric potential.

We now turn to the study of the black hole's critical behavior and thermodynamic stability. The critical point conditions are determined by:
\begin{equation}
    \left.\frac{\partial P}{\partial r_h}\right|_{T = T_c, r_h = r_c} = 0, \qquad 
    \left.\frac{\partial^2 P}{\partial r_h^2}\right|_{T = T_c, r_h = r_c} = 0,
\end{equation}
where $T_c$ and $r_c$ denote the critical temperature and horizon radius, respectively. These conditions reveal that the system exhibits critical behavior, signaling the presence of phase transitions and critical points. 
Deriving general analytical relations for the critical values of the horizon radius, pressure, and temperature in terms of the Maxwell charge is not feasible. However, by assigning specific values to the Maxwell charge \( Q \) and choosing a small perturbation coefficient \( \lambda \), we can numerically determine the critical values of the corresponding thermodynamic quantities for each charge configuration. Table~\ref{table1} presents the critical values \( r_c \), \( T_c \), and \( P_c \) for different values of \( Q \).

\begin{table}[htbp]
    \centering
    \caption{Critical values of thermodynamic quantities \( r_c \), \( T_c \), and \( P_c \) for different Maxwell charges \( Q \) with \( \lambda = 0.001 \).}
    \label{table1}
    \begin{tabularx}{0.5\textwidth}{Xcccc} 
        \toprule
        \( Q \) & \( r_c \) & \( T_c \) & \( P_c \) \\
        \midrule
        0.20 & 0.3336 & 0.3118 & 0.1729 \\
        0.30 & 0.5115 & 0.2057 & 0.0750 \\
        0.40 & 0.6869 & 0.1537 & 0.0418 \\
        0.50 & 0.8613 & 0.1228 & 0.0266 \\
        0.60 & 1.0353 & 0.1022 & 0.0185 \\
        0.70 & 1.2091 & 0.0876 & 0.0135 \\
        0.80 & 1.3827 & 0.0876 & 0.0143 \\
        0.90 & 1.5560 & 0.0681 & 0.0082 \\
        \bottomrule
    \end{tabularx}
\end{table}

\begin{figure}
\centering
\subfloat[a]{\includegraphics[width=7cm]{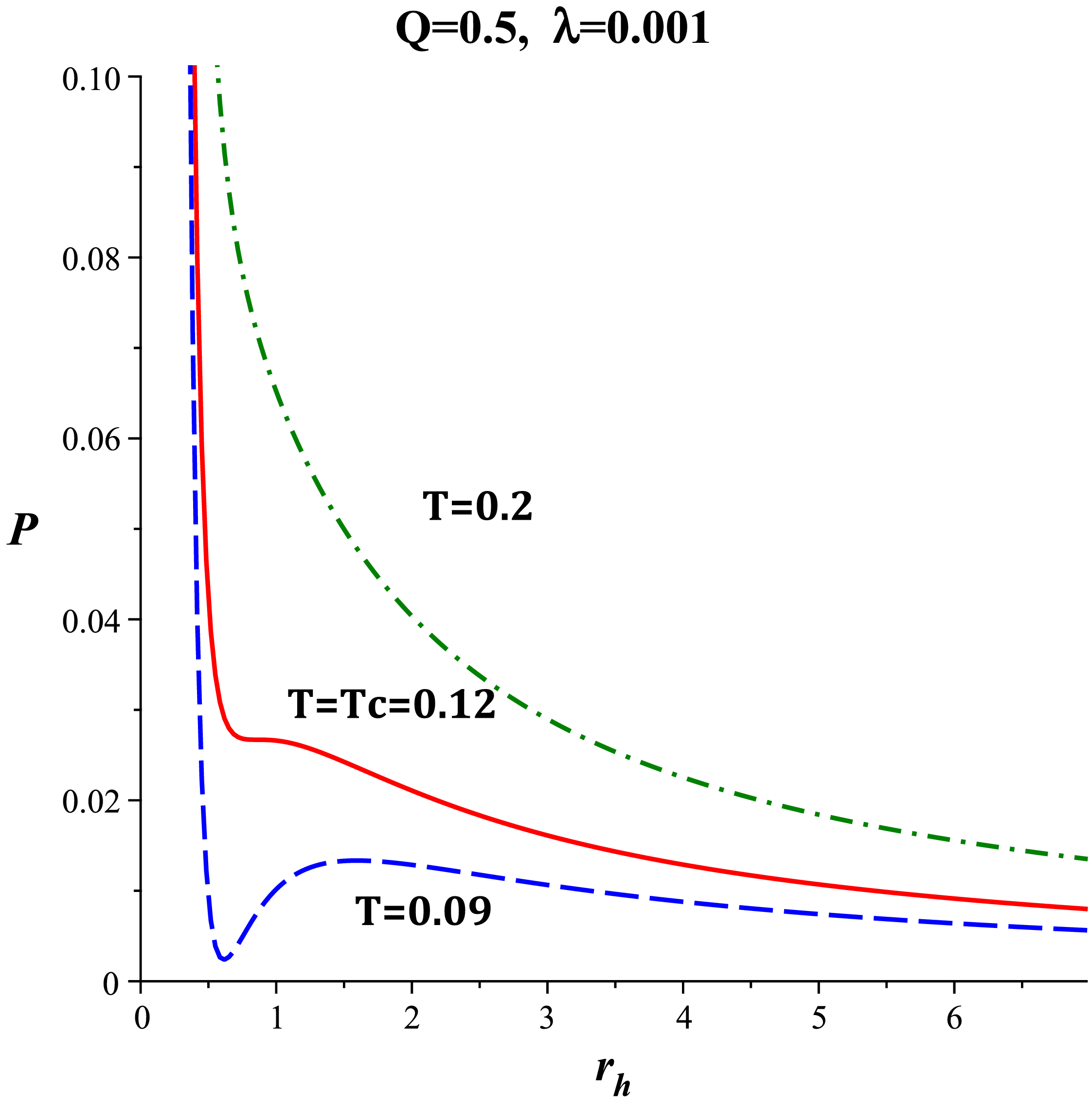}}\quad
\subfloat[b]{\includegraphics[width=7cm]{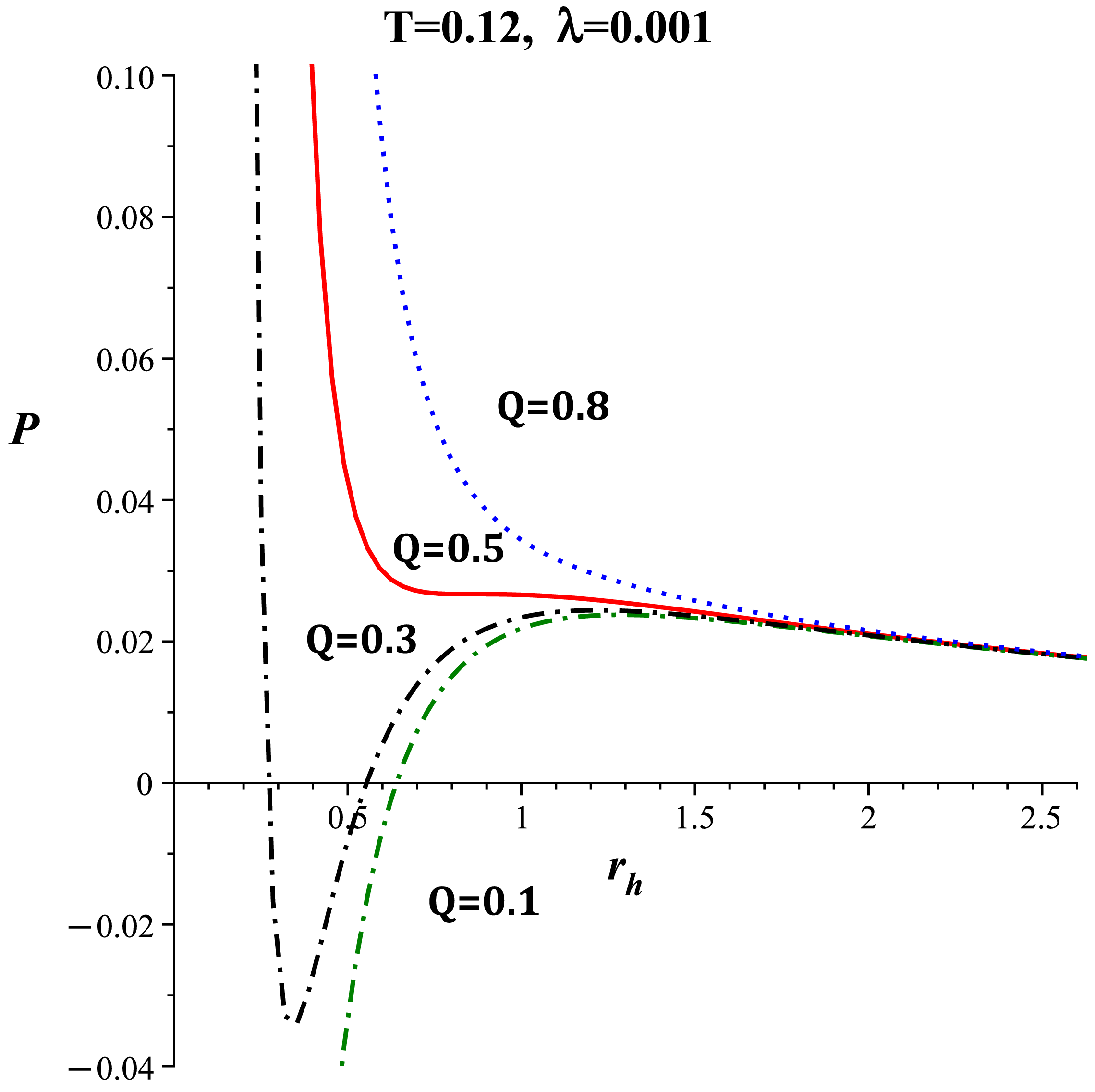}}
\caption{$P$-$r_h$ diagrams of the system: \textbf{(a)} Temperature-dependent behavior with fixed charge $Q = 0.5$; \textbf{(b)} Charge-dependent behavior ($Q$) at fixed temperature. For both cases, $\lambda = 0.001$. The black hole exhibits Van der Waals-like phase transition in general, but for $Q = 0.1$, the pressure profile closely resembles that of Hawking-Page phase transitions. \label{fig:2}}
\end{figure}

The resulting diagrams in Fig.~\ref{fig:2} reveal that the system's behavior closely resembles that of a Van der Waals fluid. Specifically, the phase transition occurs between small, intermediate, and large black hole phases, analogous to the solid-liquid-gas transition in a Van der Waals system. This behavior corresponds to a first-order phase transition, where the pressure remains constant during the phase coexistence at a fixed temperature. As the temperature increases, the system's behavior approaches that of an ideal gas, as expected.  

The right panel of Fig.~\ref{fig:2} shows that as the Maxwell charge \( Q \) decreases, the pressure diagram begins to align with the characteristics of the Hawking–Page phase transition. This is consistent with expectations, as the system tends toward a matter-free AdS spacetime in the absence of nonlinear gravitational effects and additional matter contributions.
 
\begin{figure}[H]
\centering
\subfloat[a]{\includegraphics[width=7cm]{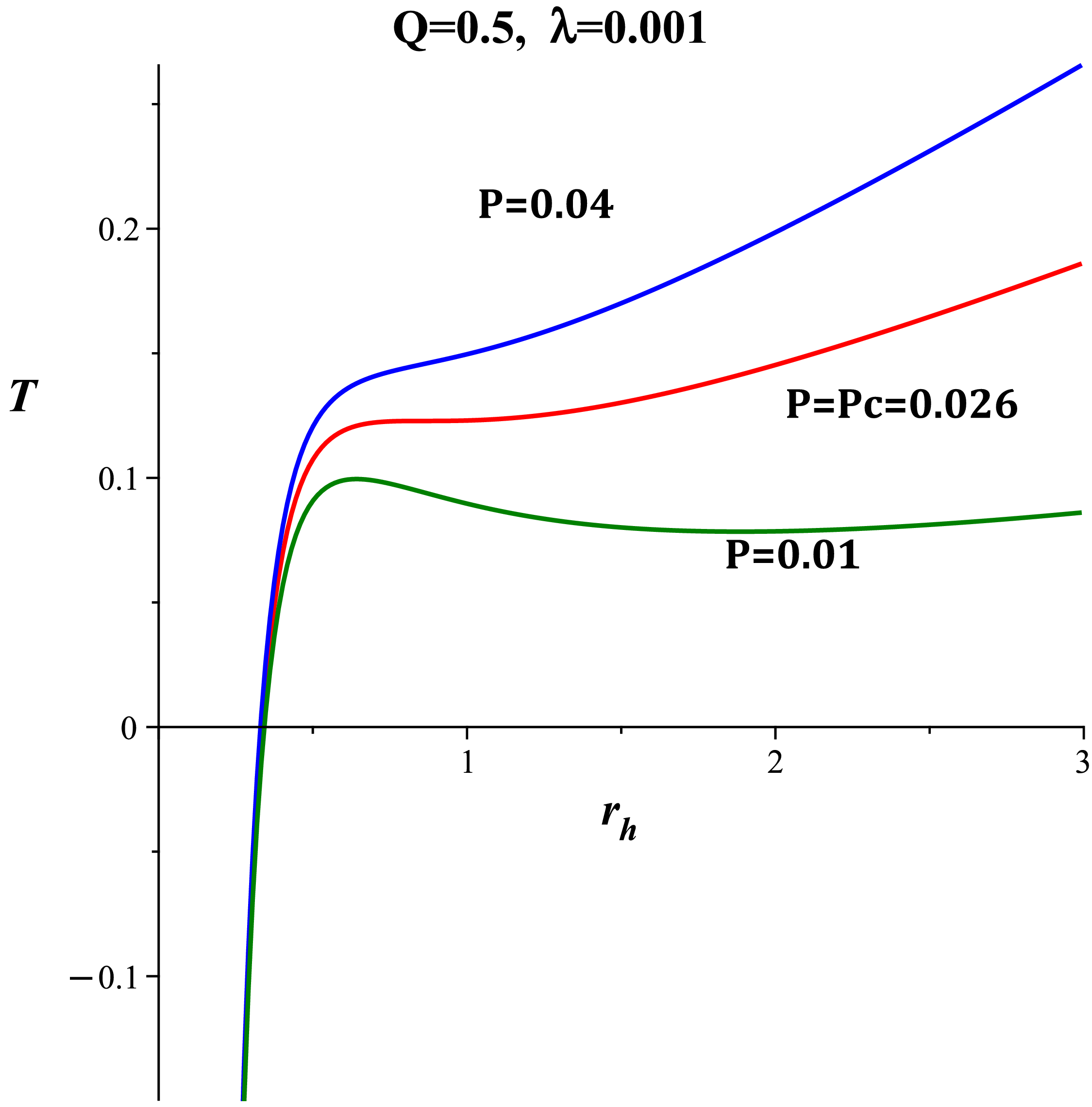}}\quad
\subfloat[b]{\includegraphics[width=7cm]{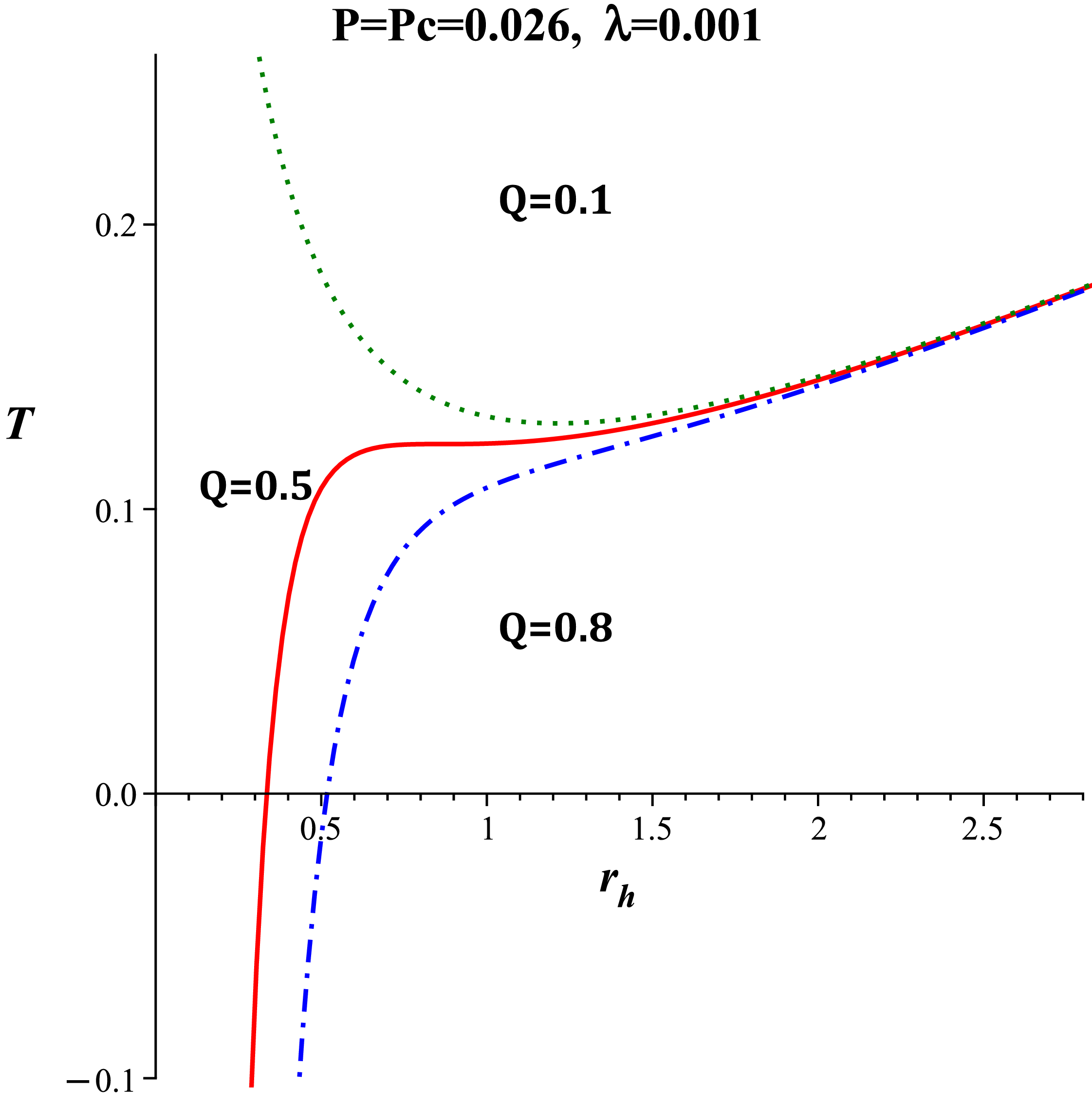}}
\caption{$T-r_{h}$ diagrams of the system: \textbf{(a)} when the system pressure changes, and \textbf{(b)} when the Maxwell charge varies. For both panels, $\lambda=0.001$. In the right panel, for $Q=0.1$, the temperature profile closely resembles that of a Hawking-Page phase transition. \label{fig:3}}
\end{figure}

The temperature behavior is depicted in Fig.~\ref{fig:3}. In the left panel, the Maxwell charge is set to $Q=0.5$, while in the right panel, the pressure is set to $P=P_c= 0.026$. In the left panel, the pressure varies. And in the right panel the Maxwell charge varies. The right panel shows that for a charge value of $Q=0.1$, the temperature diagram closely resembles that of the Hawking–Page phase transition.
It is not straightforward to analytically determine the critical value of charge at which the system's behavior transitions, for a fixed \(\lambda\), from a Hawking--Page phase transition to a Van der Waals--like phase transition. Instead, we must rely on numerical analysis. For \(Q = 0\), the system exhibits Hawking-Page-like behavior, but as the charge increases, it gradually transitions to Van der Waals fluid behavior. However, a sharp distinction between the two regimes is not clearly observable. For example, we observed that for \(Q = 0.1\) and \(\lambda = 0.001\), the system exhibits characteristics similar to Hawking-Page phase transition, while for \(Q = 0.5\), it displays Van der Waals-like characteristics. Through numerical investigations, we found that for \(\lambda = 0.001\), the Hawking--Page behavior is dominant for approximately \(Q < 0.15\). One possible approach is to use the critical point condition for temperature or pressure to determine whether it is satisfied for a given value of \( Q \). For some values of \( Q \), the critical condition yields no real roots for the black hole horizon---the solutions are imaginary. However, for other values, real roots can be found that satisfy the critical point condition. The existence of such real roots is a characteristic feature of Van der Waals--type fluids. Note that we select only those real roots which lead to a positive critical pressure, as required in asymptotically AdS space.
 
Phase transitions in canonical ensemble can be classified using the Helmholtz free energy by examining how the free energy and its derivatives change during the transition. First-order phase transitions are characterized by a discontinuity in the first derivative of the free energy with respect to some thermodynamic variable (e.g., temperature or pressure). At the transition point, the system absorbs or releases a latent heat. The Helmholtz free energy changes smoothly at the transition point, but its slope (the first derivative) is discontinuous. Above the transition point, it becomes smoother. Typically, for a Van der Waals fluid, below the critical pressure or temperature, a swallowtail-like shape is observed.\par 
Second-order (or Continuous) phase transitions are characterized by a continuous first derivative of the Helmholtz free energy, but a discontinuity in the second derivative. There is no latent heat associated with second-order transitions. Examples include the transition from a ferromagnetic to a paramagnetic state in certain materials at the Curie point. The Helmholtz free energy  and its first derivative are continuous, but the second derivative (e.g., heat capacity, magnetic susceptibility) shows a divergence or discontinuity. \par
\begin{figure}[H]
\centering
\subfloat[a]{\includegraphics[width=5.5cm]{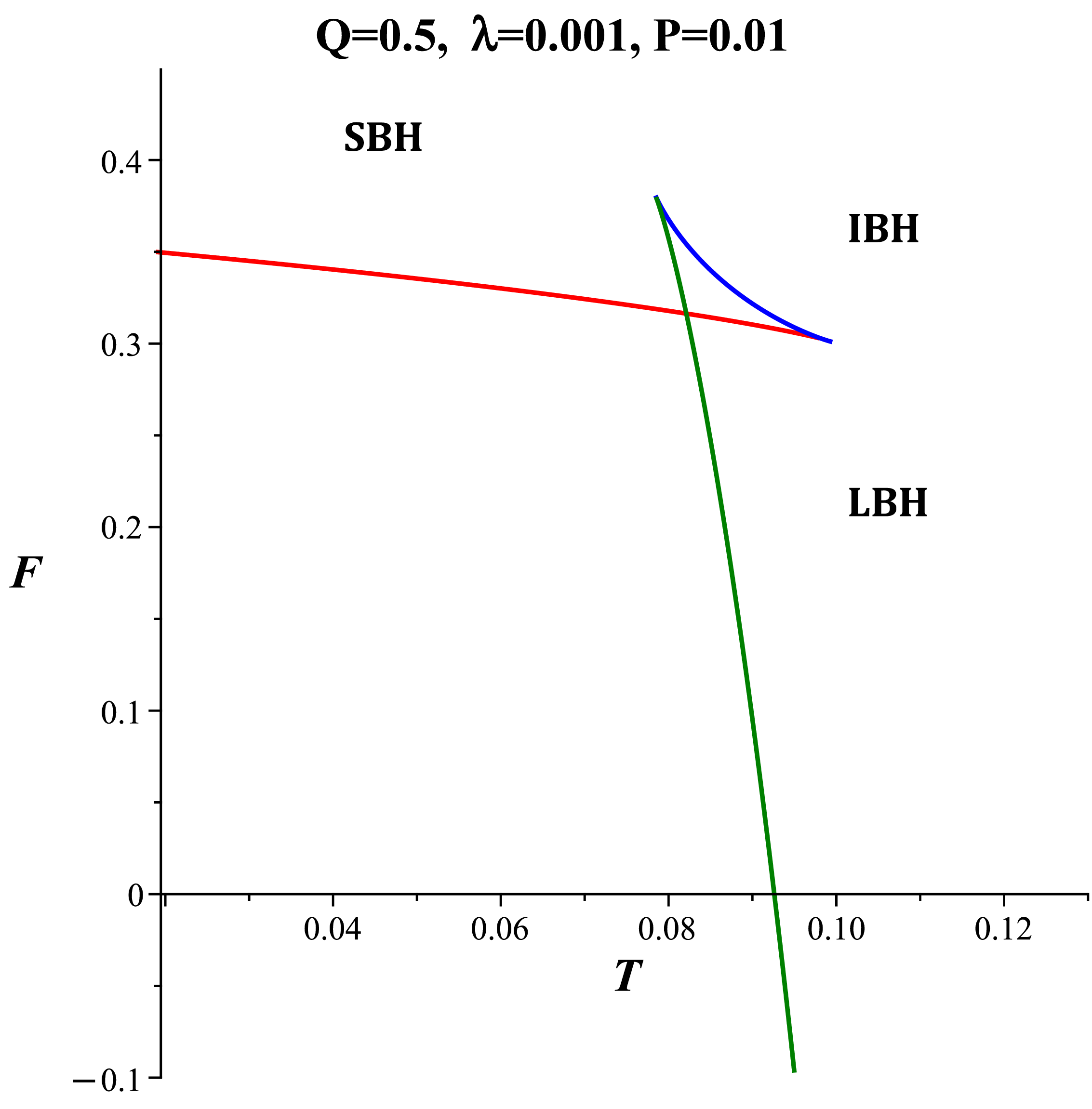}}\quad
\subfloat[b]{\includegraphics[width=5.5cm]{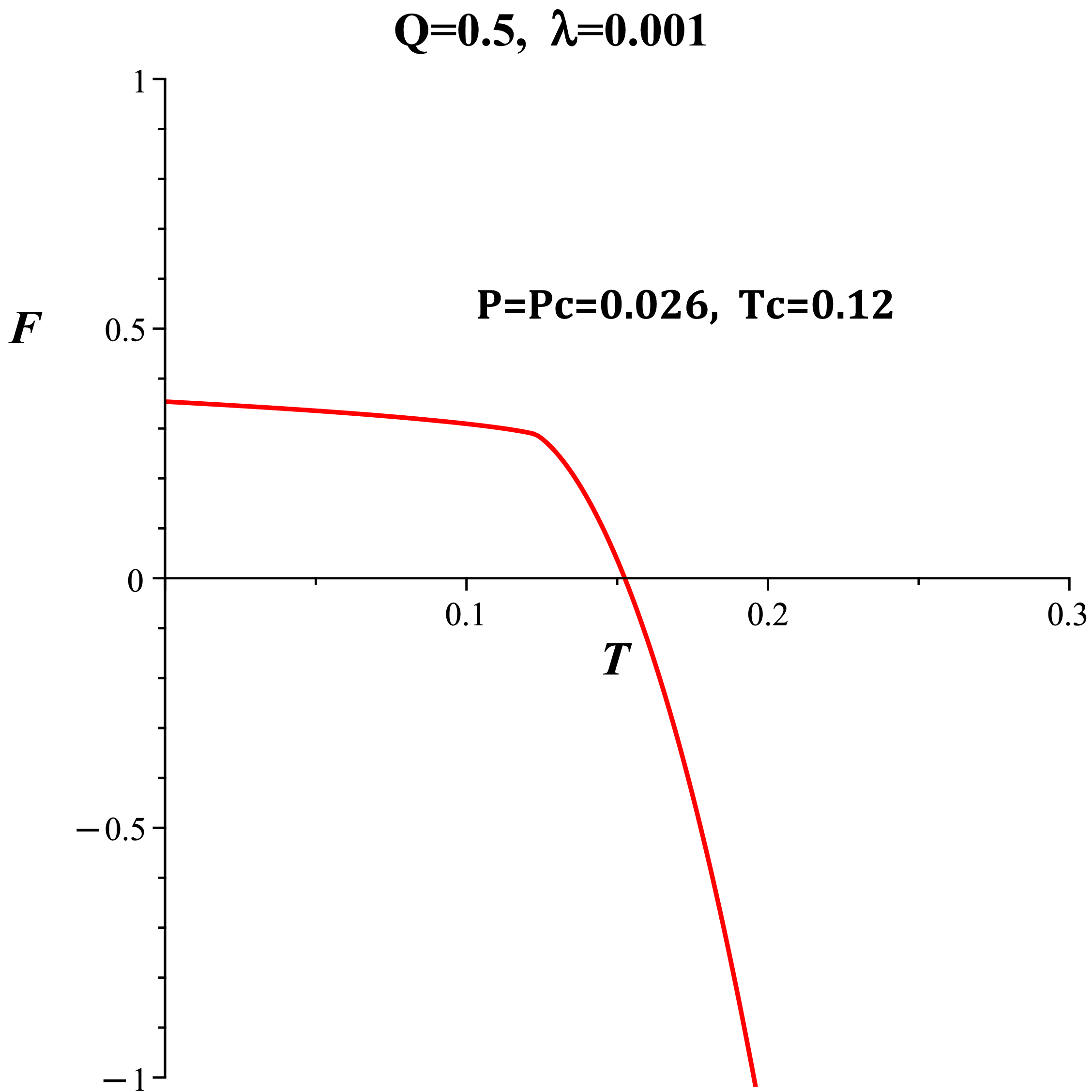}}\quad
\subfloat[c]{\includegraphics[width=5.5cm]{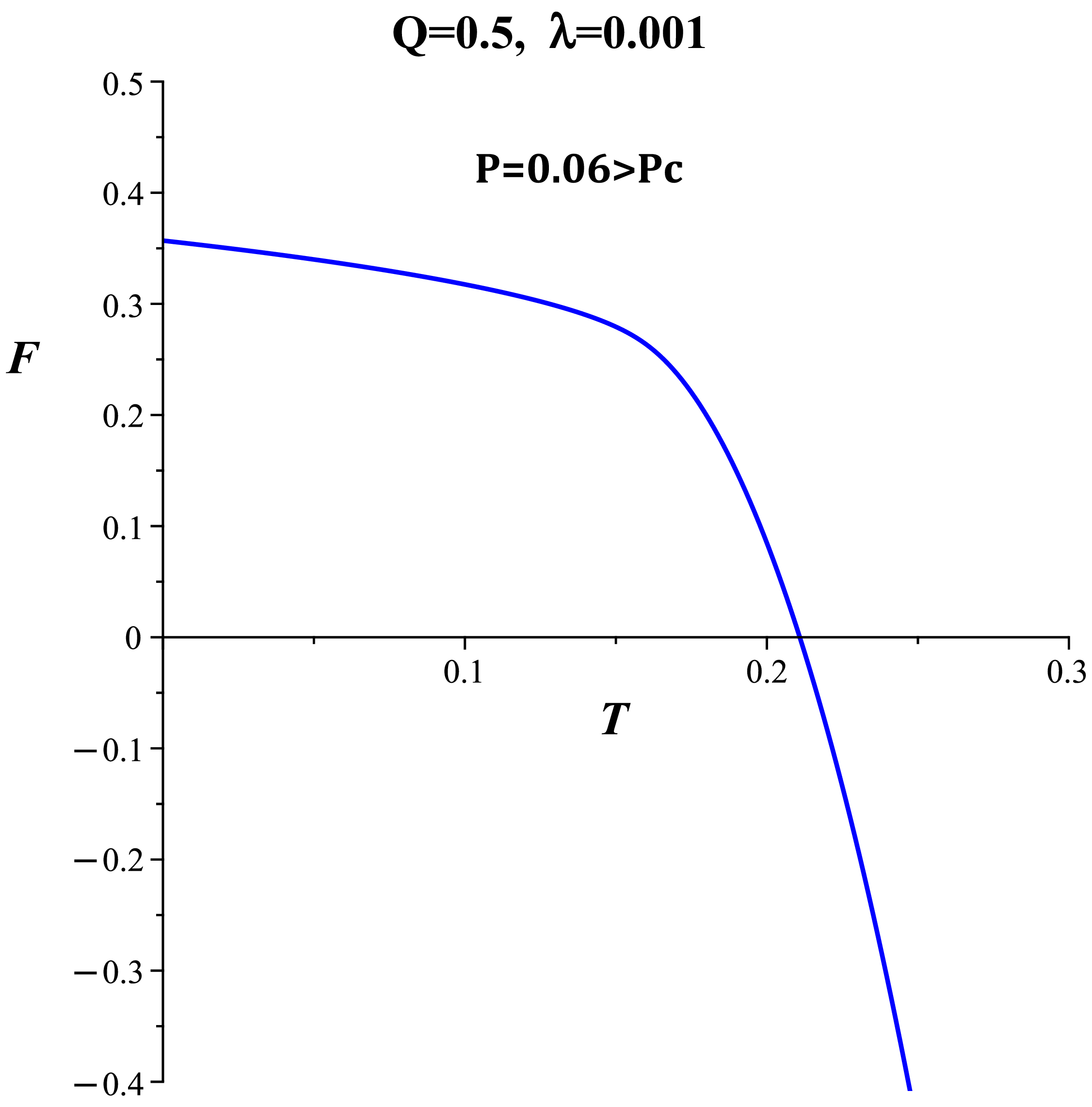}}
\caption{$F-T$ diagrams of the system: \textbf{(a)} with a pressure of $P = 0.01$ (below the critical pressure), \textbf{(b)} at the critical pressure $P_c = 0.026$, and \textbf{(c)} with a pressure of $P = 0.06$ (above the critical pressure). For all diagrams, $\lambda = 0.001$. The swallowtail shape in the left panel is characteristic of Van der Waals fluids. \label{fig:4}}
\end{figure}

 
The Helmholtz free energy as a function of the black hole's horizon, up to the first order in $\lambda$ is as follows:
\begin{equation}\label{GR}
F=H-T S = \frac{r_{h}}{4}-\frac{2 r_{h}^{3} \pi  P}{3}+\frac{3 Q^{2}}{8 r_{h}}+\left(\frac{4 Q^{2} \pi  P}{r_{h}}-\frac{Q^{2}}{2 r_{h}^{3}}-\frac{3 Q^{4}}{10 r_{h}^{5}}\right) \lambda
\end{equation}
By utilizing the Hawking temperature relation, the equation (\ref{GR}) can be reformulated in terms of the black hole's pressure and temperature. This reformulation facilitates the creation of $F$-$T$ diagrams. The left panel of Fig.~\ref{fig:4} shows the Helmholtz free energy as a function of temperature for a pressure below the critical pressure. A swallowtail shape is observed in the figure. The red, blue, and green colors correspond to the small, intermediate, and large black holes, respectively.
The middle panel shows the Helmholtz function corresponding to the critical pressure. A kink is observed in the diagram, indicating that the first derivative of the free energy at the critical temperature, which is proportional to the entropy, is not continuous. 
The right panel of the figure shows the smooth behavior of the Helmholtz function for pressures above the critical pressure.

\begin{figure}
\centering
\subfloat[a]{\includegraphics[width=5.5cm]{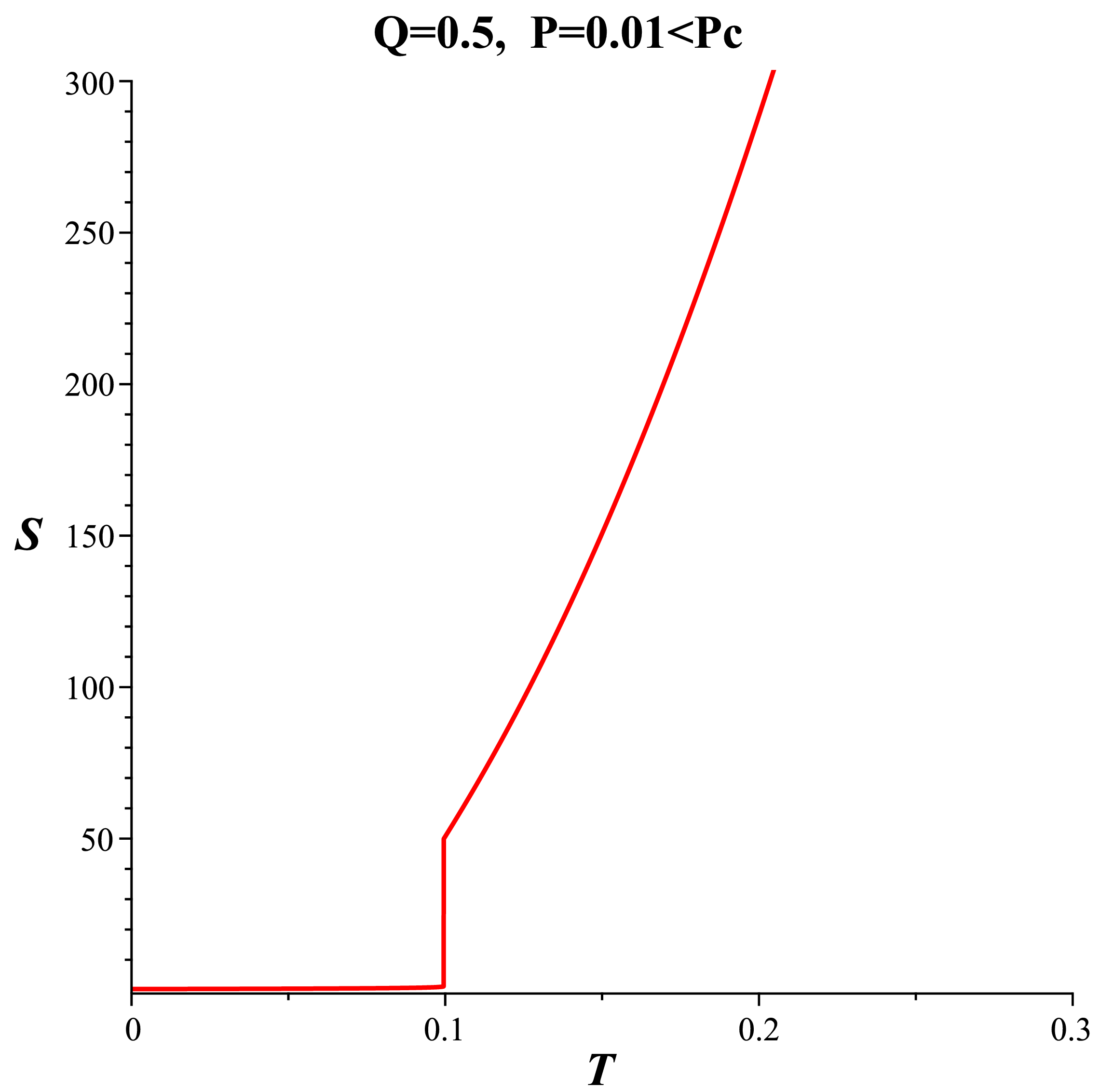}}\quad
\subfloat[b]{\includegraphics[width=5.5cm]{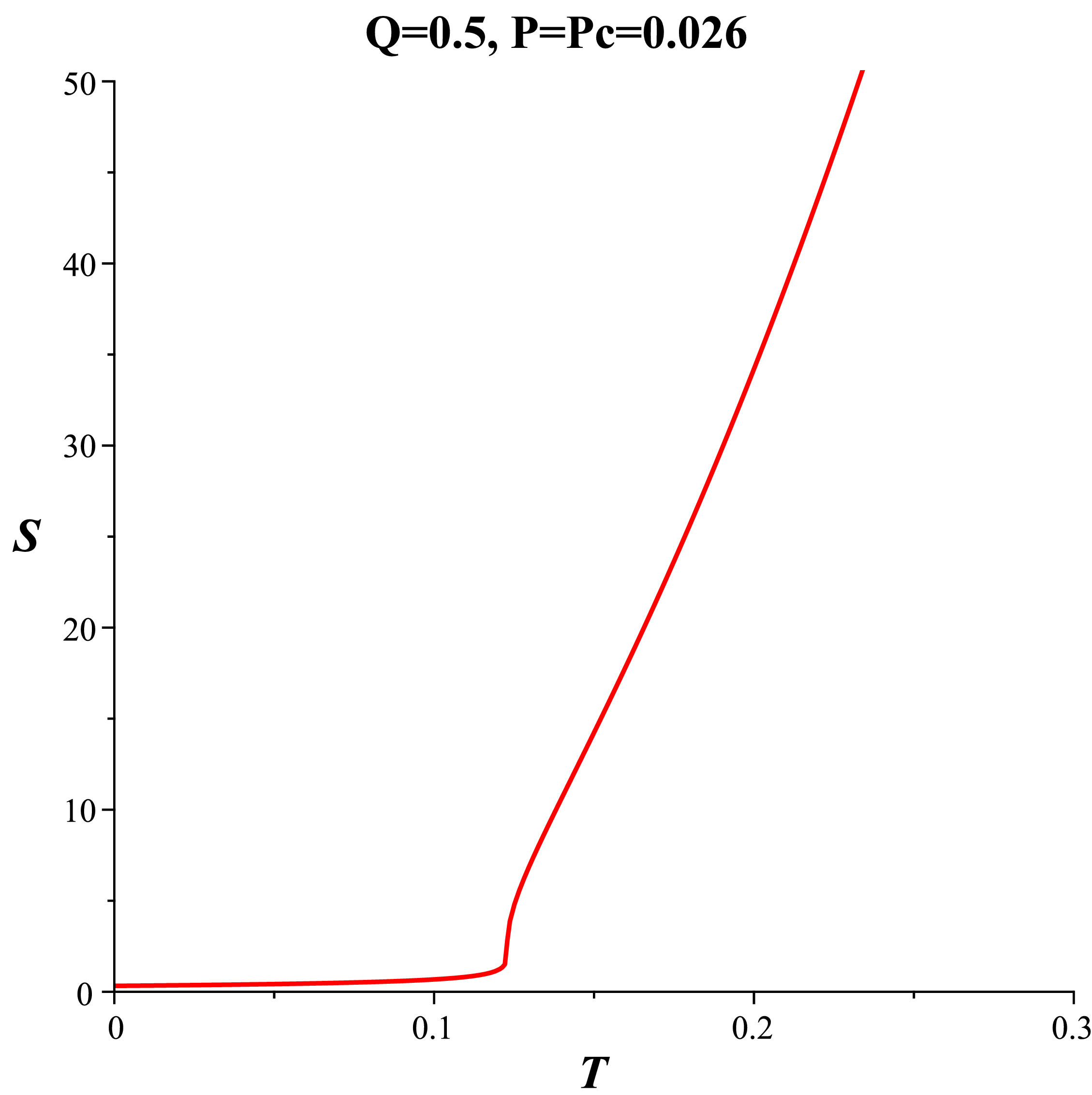}}\quad
\subfloat[c]{\includegraphics[width=5.5cm]{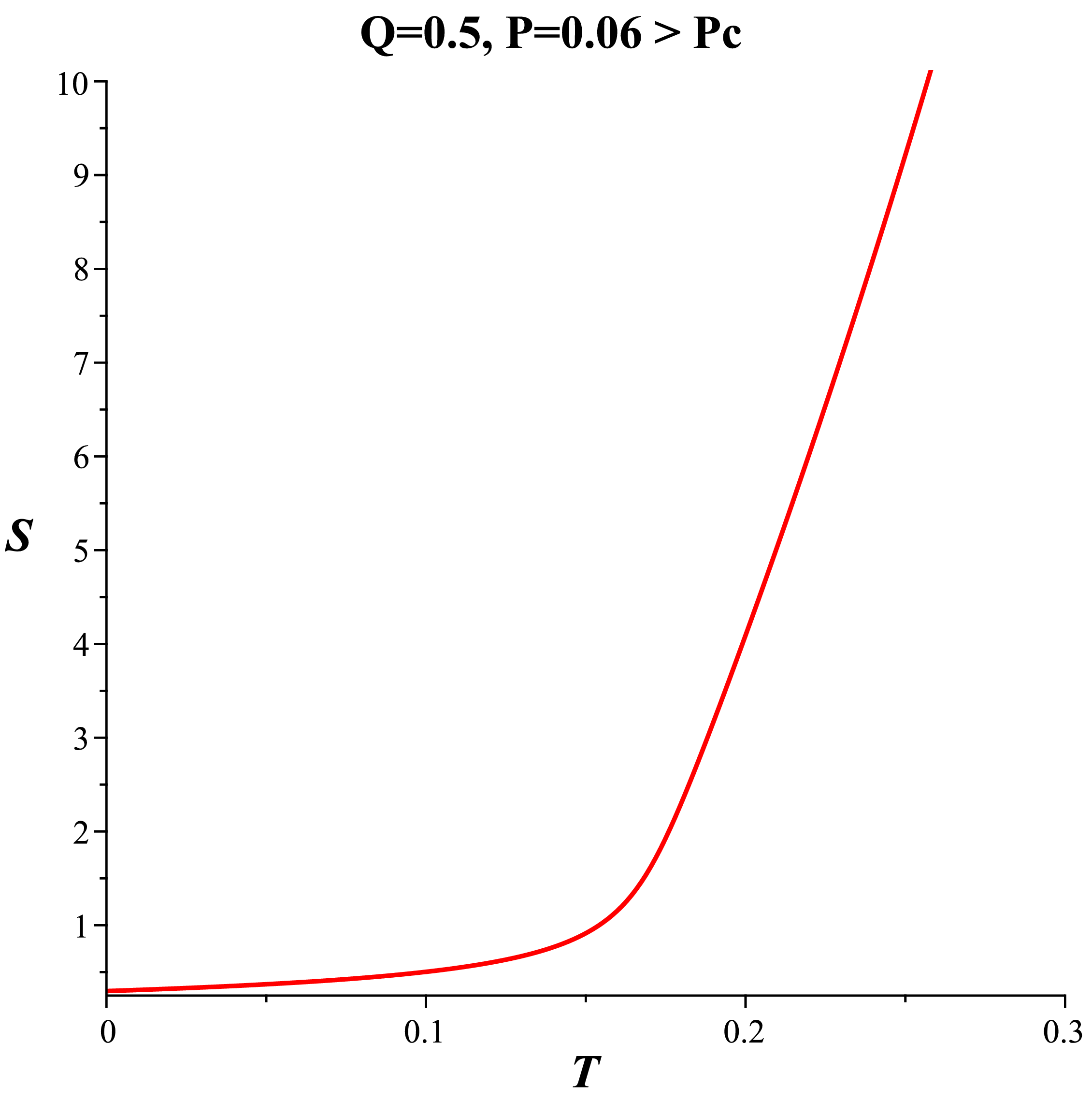}}
\caption{$S-T$ diagrams of the system: \textbf{(a)} with a pressure of $P = 0.01$, which is less than the critical pressure, \textbf{(b)} at the critical pressure of $P = P_c = 0.026$, and \textbf{(c)} with a pressure of $P = 0.06$, which is higher than the critical pressure. For all diagrams, $\lambda=0.001$. A jump is seen in the left panel which is characteristic of a first order phase transions. \label{fig:5}}
\end{figure}

In a typical Van der Waals-like phase transition, a discontinuity in entropy should be observed. As shown in Fig.~\ref{fig:5}, for pressures below the critical pressure, a clear entropy jump appears in the left panel, indicating the presence of latent heat. This behavior is characteristic of first-order (discontinuous) phase transitions. The middle panel displays critical behavior at the critical pressure, marking the boundary between discontinuous and smooth behaviors. Finally, the right panel exhibits a smooth behavior for pressures above the critical pressure.

\begin{figure}[H]
\centering
\subfloat[a]{\includegraphics[width=5.5cm]{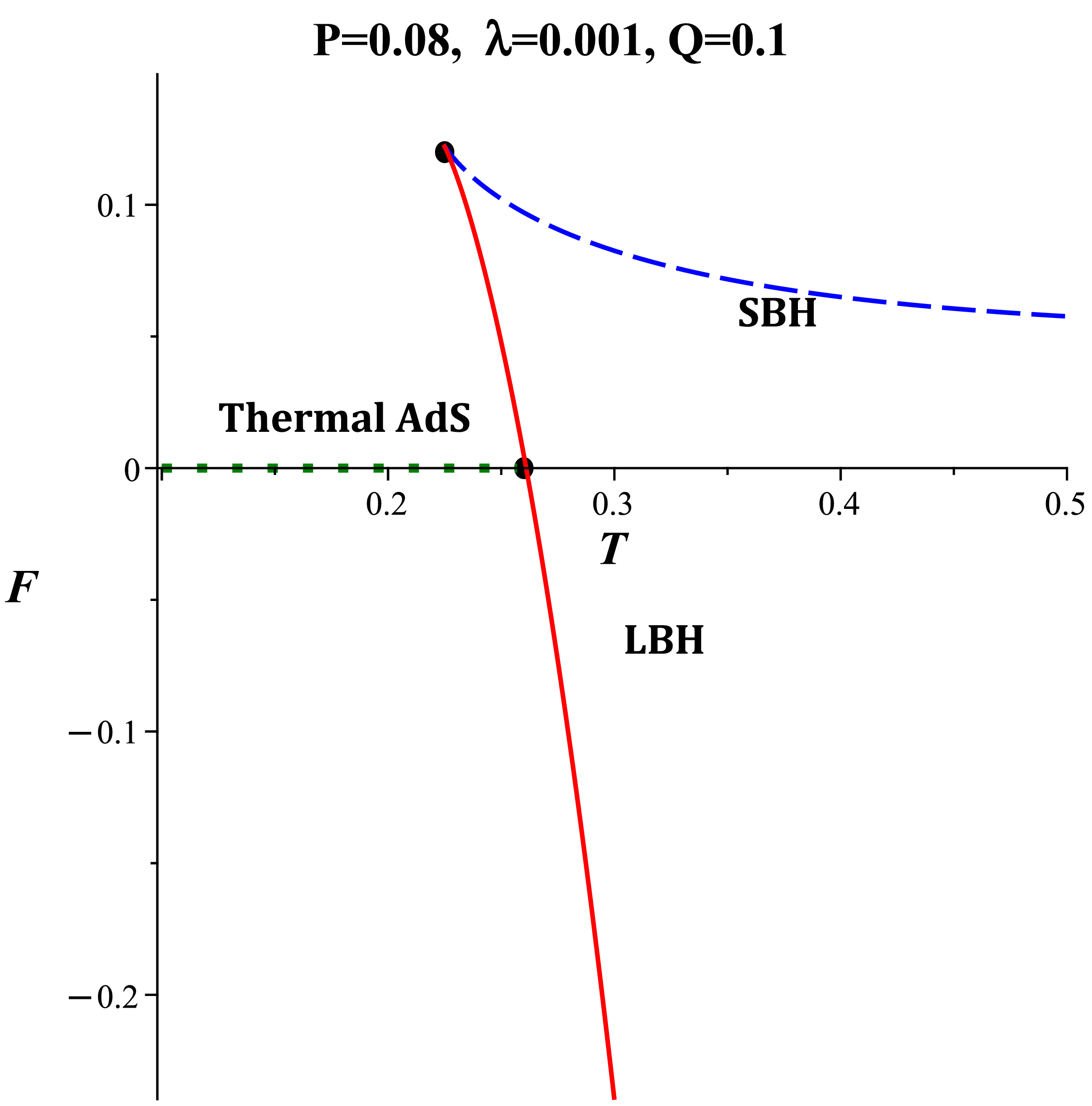}}\quad
\subfloat[b]{\includegraphics[width=5.5cm]{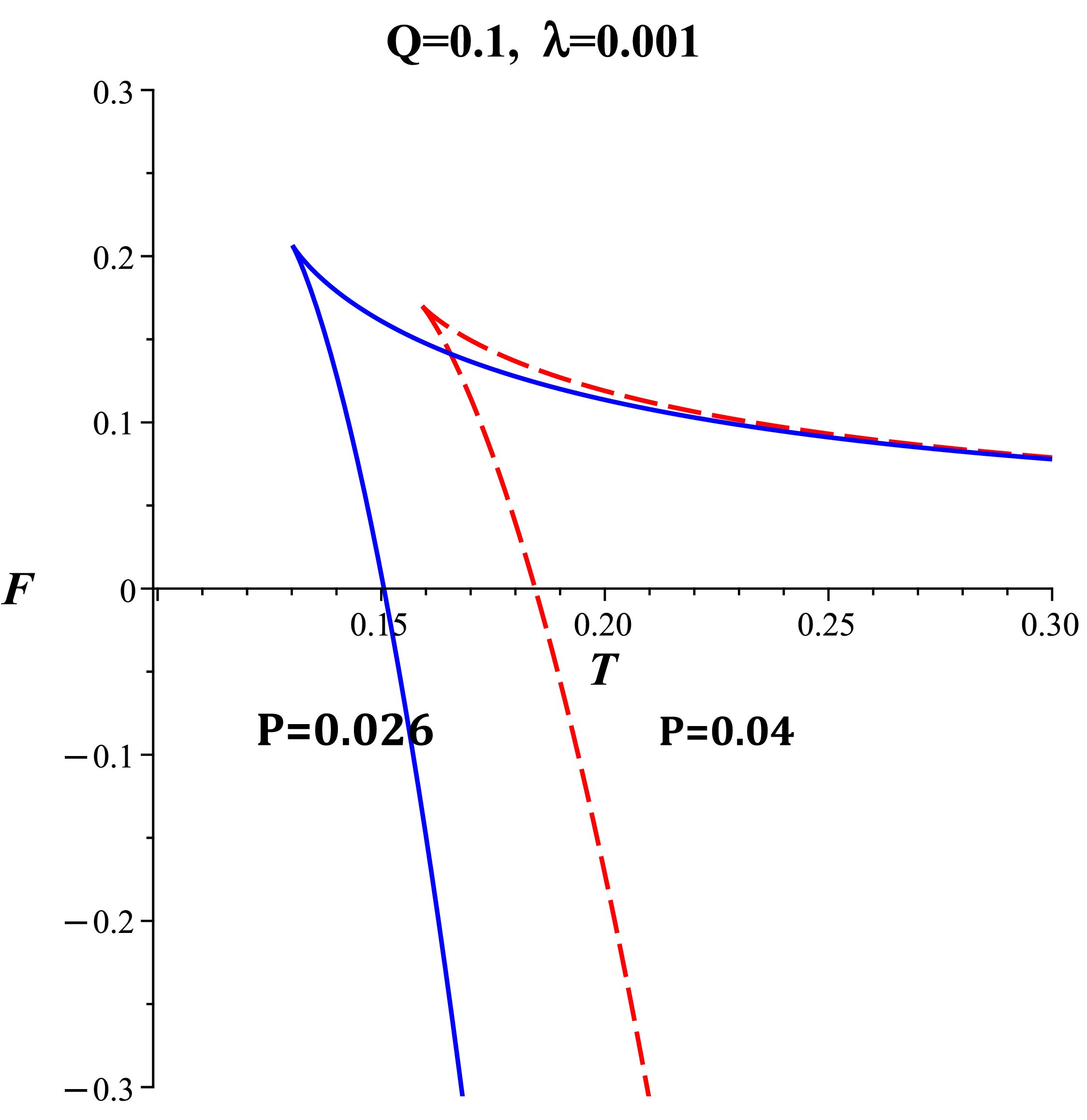}}\quad
\subfloat[c]{\includegraphics[width=5.5cm]{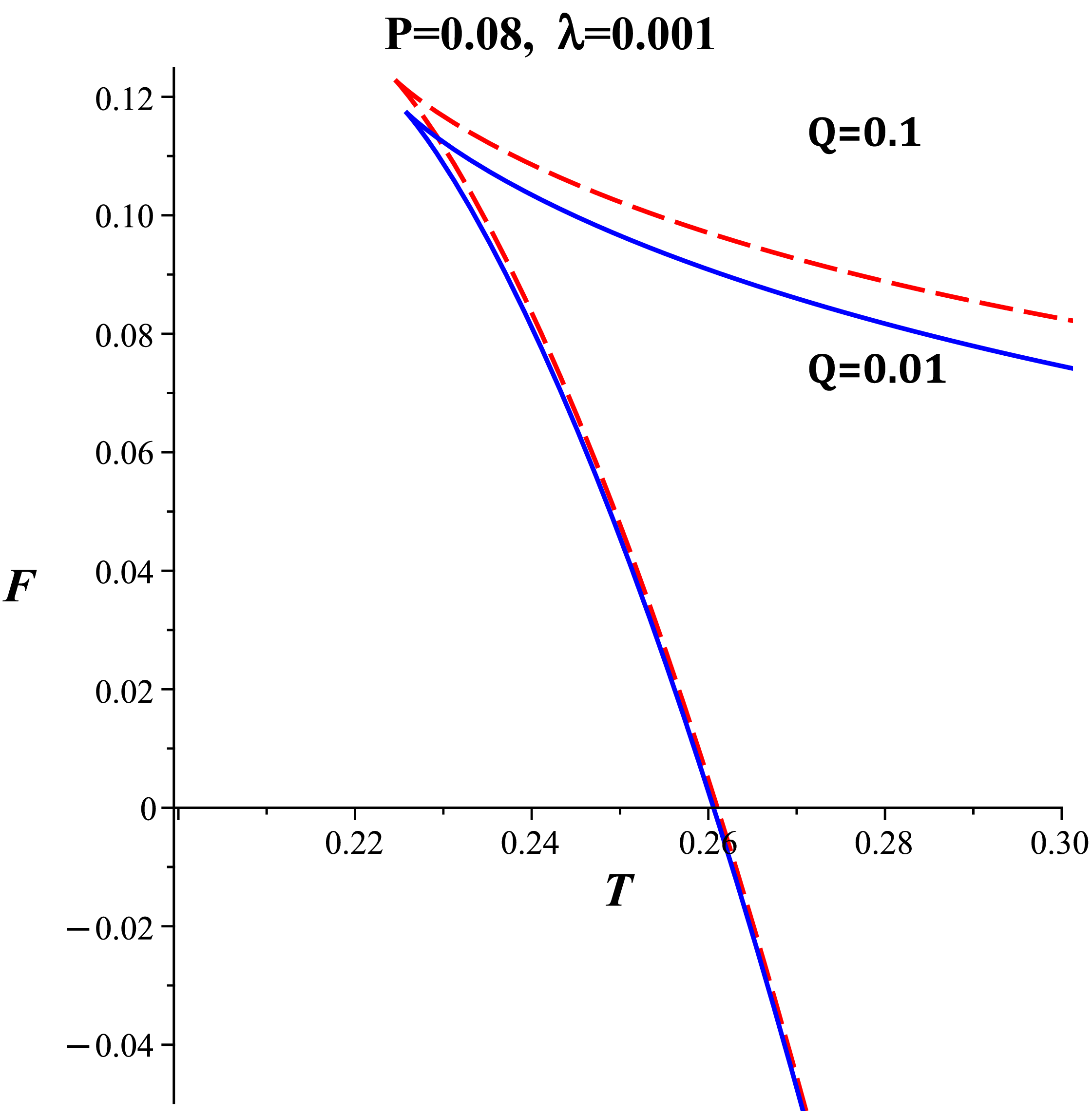}}
\caption{$F$-$T$ diagrams of the black hole: 
\textbf{(a)} Different regions of the Helmholtz free energy in a Hawking-Page-like phase transition, \textbf{(b)} Changes in the pressure of the black hole with $Q=0.1$, \textbf{(c)} Variations in $Q$ with a fixed pressure of $P=0.08$. For all panels, $\lambda=0.001$ is used as a small perturbation coefficient. \label{fig:6}}
\end{figure}

If we decrease the value of the Maxwell charge, as we expect a Hawking-Page like behavior is seen for the Helmholtz free energy.
In the regime where the black hole phase is dominant, the upper branch of the Helmholtz function represents small black holes, while the lower branches represent larger black holes. At the point where $F = 0$, a transition between the thermal AdS space and stable large black holes is possible. This condition, $F = 0$, can be utilized to determine the transition radius $r_{HP}$ and the Hawking-Page temperature $T_{HP}$ between the thermal AdS space and the large black hole state. Refer to the left panel of Fig.~\ref{fig:6}.
In the middle panel of Fig.~\ref{fig:6}, the pressure of the black hole changes, while in the right panel, the Maxwell charge changes.
Specifically, the Helmholtz free energy shows a distinct minimum, where LBH and SBH curves meet each other, that corresponds to the critical point of the phase transition. Below this critical point, the system remains in a stable thermal radiation phase, whereas above it, the black hole phase becomes dominant. See Fig.~\ref{fig:6}. 

The local stability of the black hole can be checked by the behavior of the heat capacity of the black hole. In thermodynamic terms, a black hole is locally stable if its heat capacity at constant pressure is positive. The positivity of the heat capacity ensures that small fluctuations in the black hole's temperature will not lead to runaway growth in either energy absorption or emission. 
Usually, the roots of the heat capacity at constant pressure for an AdS black hole indicate the points where the heat capacity changes sign. These roots are significant because they mark the transitions between stable and unstable phases of the black hole. 
The divergence points of the heat capacity of an AdS black hole indicate critical points where the black hole undergoes phase transitions. These points are essential for understanding the thermodynamic stability and phase structure of the black hole. The divergence points in specific heat capacity often signal second-order phase transitions. Because, at a second-order phase transition, the first derivative of the free energy is continuous, but the second derivative (such as specific heat) diverges. However, the existence of divergence points alone does not confirm a second-order phase transition. There are additional ways to confirm a second-order phase transition beyond just looking at the divergence points of heat capacity. The second-order phase transitions often involve a continuous change in an order parameter, which is a quantity that characterizes the phase of the system (e.g., magnetization in ferromagnetic transitions). Another way is investigating the critical exponents. The critical exponents describe the behavior of physical quantities near the critical point. In second-order phase transitions, these exponents follow specific scaling laws.  The investigation of susceptibility also helps in this regard. The susceptibility, measures the response of the system to an external field, often diverges at second-order transitions\cite{dombgreen1972}. The theoretical approaches are often useful. For example \textit{renormalization group theory} helps to analyze critical behavior by studying how the system's properties change with scale, confirming the nature of the phase transition\cite{wamer2018}.\par
\begin{figure}
\centering
\subfloat[a]{\includegraphics[width=5.5cm]{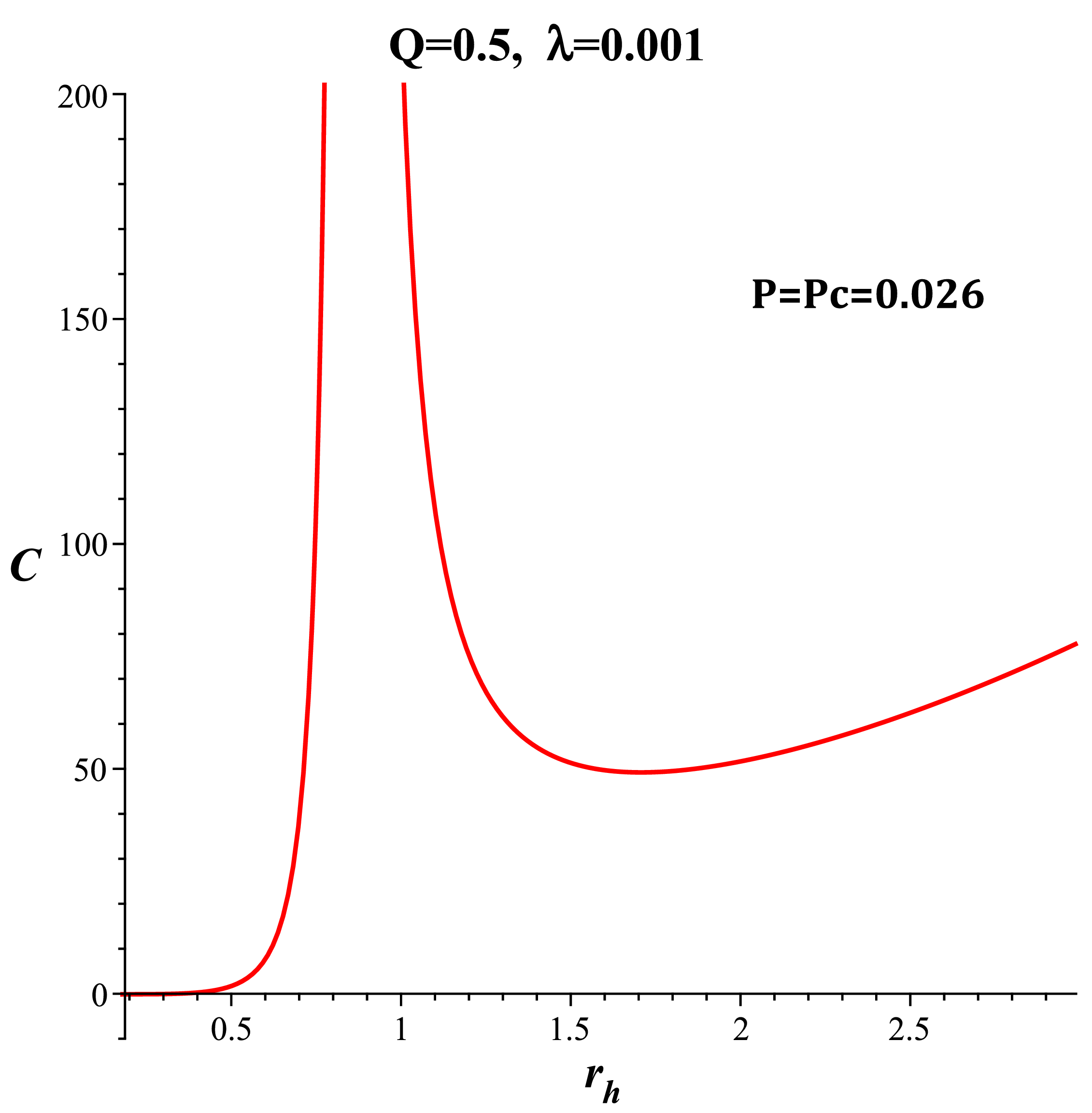}}\quad
\subfloat[b]{\includegraphics[width=5.5cm]{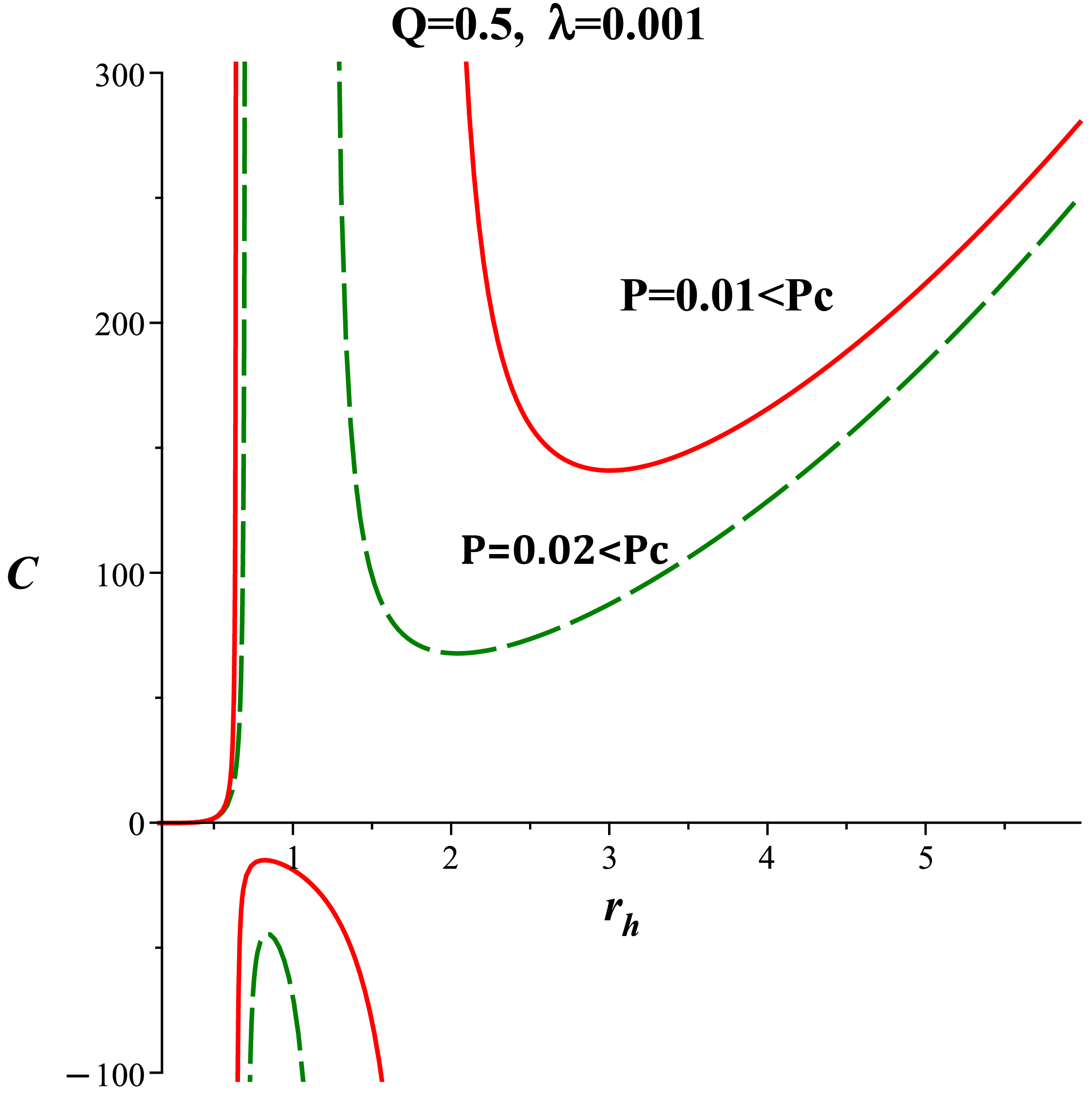}}\quad
\subfloat[c]{\includegraphics[width=5.5cm]{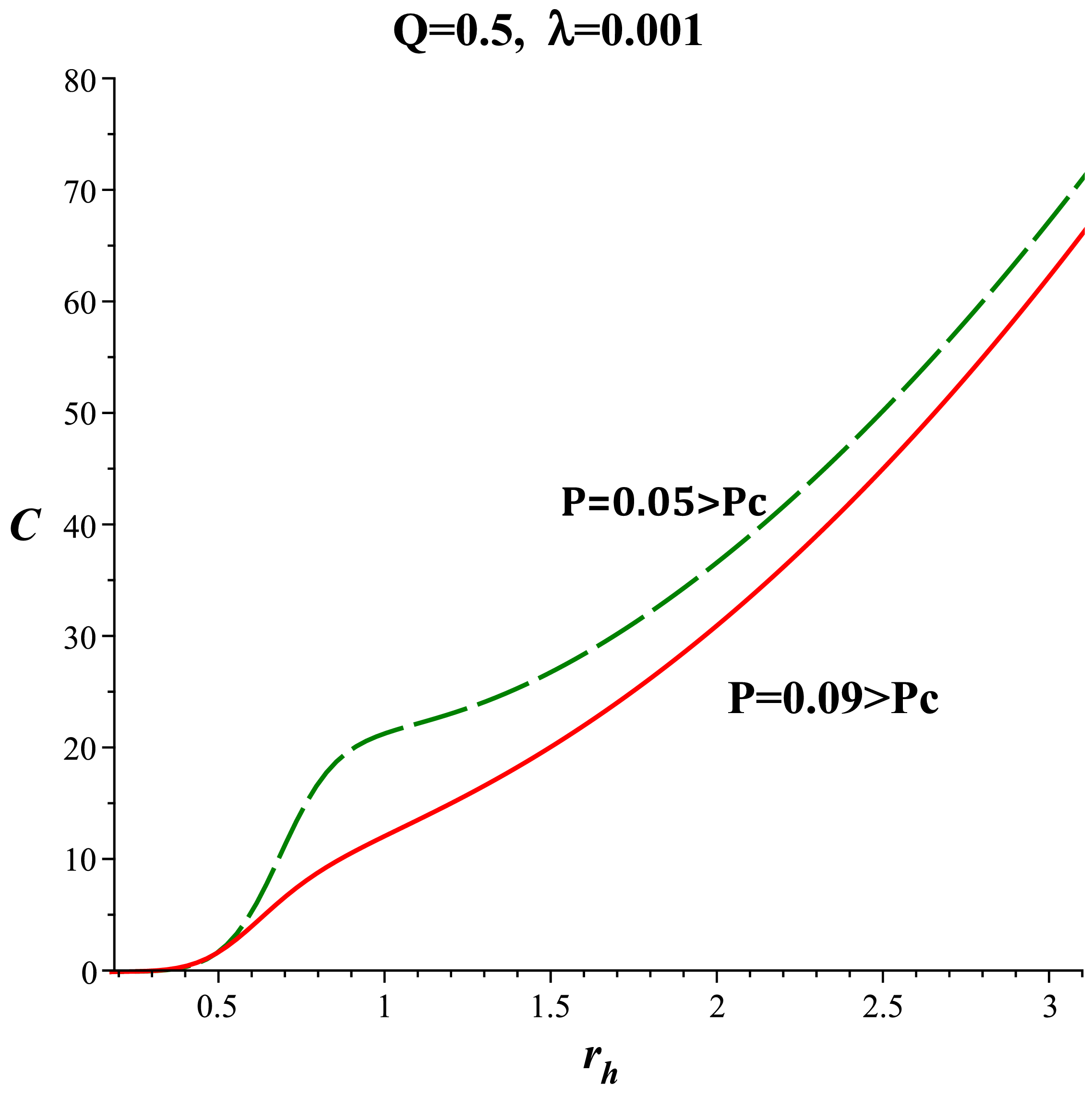}}
\caption{$C-r_h$ diagrams of the system: \textbf{(a)} at the critical pressure, $P = P_c = 0.026$, \textbf{(b)} at pressures less than the critical pressure, and \textbf{(c)} at pressures higher than the critical pressure. For all diagrams, $\lambda = 0.001$ and $Q = 0.5$. At the critical pressure, the intermediate region disappears. 
\label{fig:7}}
\end{figure}

 
The form of the heat capacity at constant pressure as a function of the black hole's horizon is as follows:
\begin{equation}
C_P= \left[\frac{\partial H}{\partial r_h} \left(\frac{\partial T}{\partial r_h}\right)^{-1}\right]_{P} = \frac{\frac{1}{2}+4\pi P r_{h}^{2}-\frac{Q^{2}}{4 r_{h}^{2}}+\left(\frac{8 Q^{2} \pi  P}{r_{h}^{2}}-\frac{Q^{4}}{r_{h}^{6}}+\frac{3 Q^{2}}{r_{h}^{4}}\right) \lambda}{2 P+\frac{3 Q^{2}}{8 \pi  r_{h}^{4}}-\frac{1}{4 \pi  r_{h}^{2}}+\left(\frac{12 Q^{2} P}{r_{h}^{4}}-\frac{5 Q^{2}}{2 \pi  r_{h}^{6}}\right) \lambda}.
\end{equation}
Fig.~\ref{fig:7}, represents the behavior of the heat capacity as a function of  black hole's horizon for a fixed charge of $Q=0.5$ and different values of pressure. 
At the critical pressure, only stable and unstable regions are present (left panel). When the pressure is below the critical pressure, an  unstable intermediate region also emerges (middle panel). Conversely, for pressures exceeding the critical pressure, the stable region becomes predominant, and the unstable region vanishes (right panel).

\begin{figure}
\centering
\subfloat[a]{\includegraphics[width=7cm]{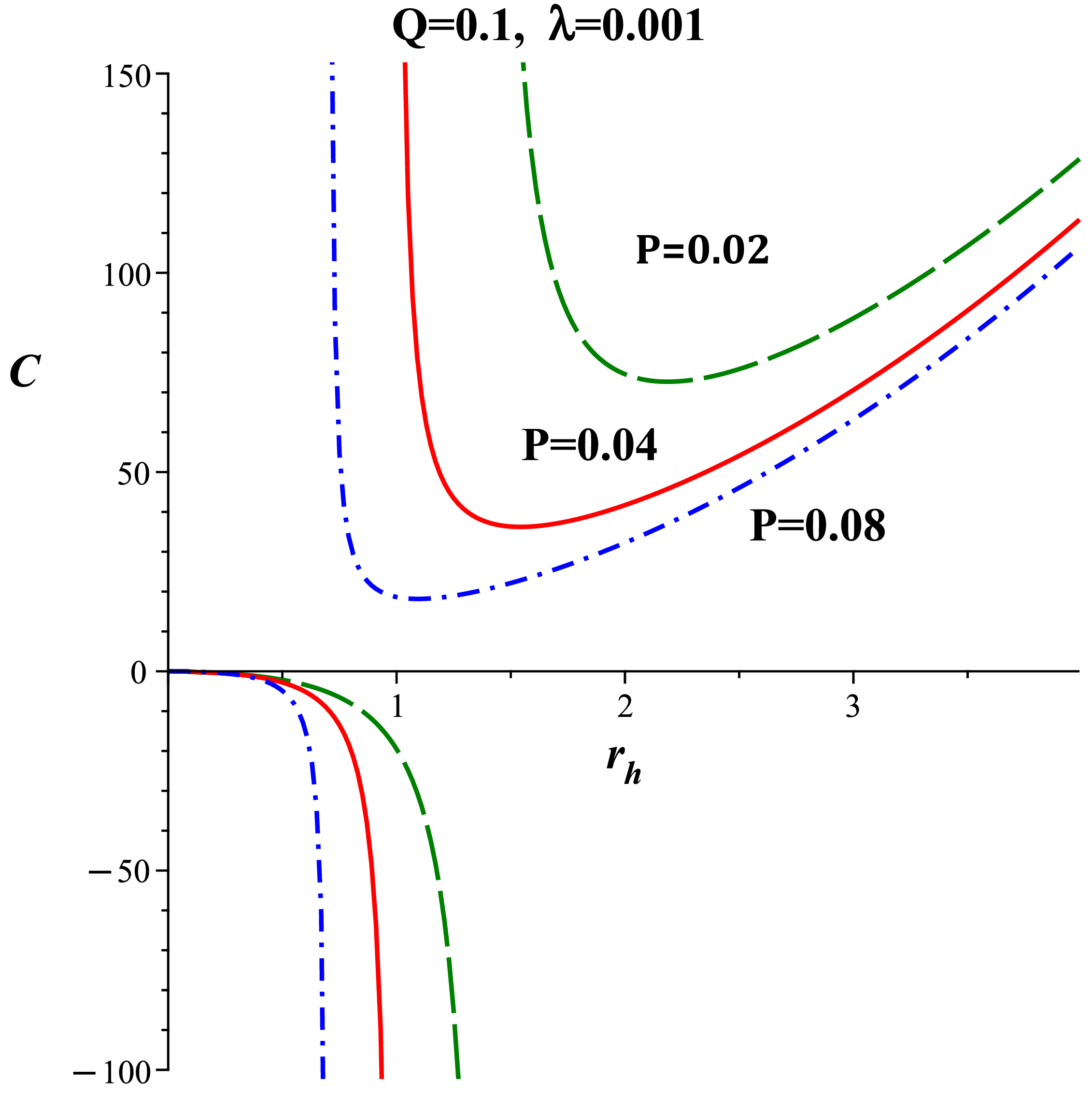}}\quad
\subfloat[b]{\includegraphics[width=7cm]{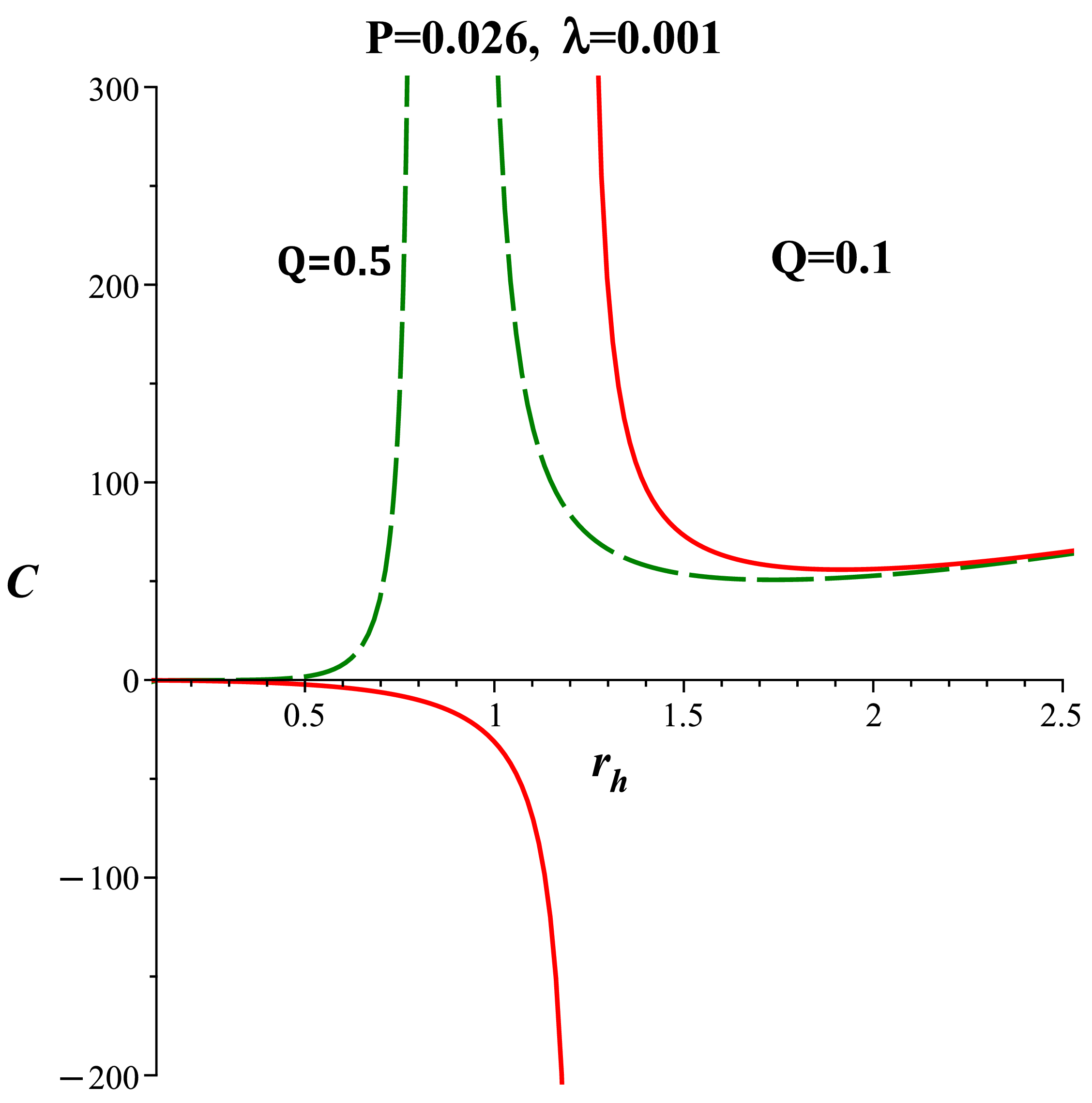}}
\caption{$C$-$r_h$ diagrams of the black hole for fixed charge $Q = 0.1$:  
\textbf{(a)} Dependence of heat capacity ($C$) on the horizon radius ($r_h$) under varying pressure at $Q = 0.1$,  
\textbf{(b)} Behavior at fixed pressure $P = 0.026$ with varying charge $Q$. For comparison, curves for both $Q = 0.5$ and $Q = 0.1$ are plotted together. The perturbation coefficient is fixed at $\lambda = 0.001$ in both cases.  
\label{fig:8}}
\end{figure}
The heat capacity analysis of the black hole for showing the low charge regime, illustrated in Fig.~\ref{fig:8}, reveals two distinct phases: unstable and stable. These phases correspond to the transition between small and large black holes observed in the Hawking-Page phase transition. Increasing the pressure causes the phase transition to occur at smaller horizon radii, and conversely, decreasing the pressure shifts the transition to larger horizons. In a low charge regime the behavior of heat capacity is close to the Hawking-Page phase transition.

\section{Grand Canonical Ensemble Analysis: Thermodynamics and Stability}\label{sec4}
In this section, we explore the thermodynamics of black holes within the framework of the grand canonical ensemble, where the Gibbs free energy serves as the corresponding thermodynamic potential. Within this ensemble, the Maxwell charge is allowed to vary while its associated potential remains fixed. 
To proceed, it is essential to express the necessary thermodynamic quantities in terms of the conjugate potential, $\Phi$ to the Maxwell charge which has been obtained in relation (\ref{fi}). The Maxwell charge can be obtained from the equation (\ref{fi}) and inserted into the necessary thermodynamic relations.  \par
The enthalpy of the black hole, to the first order in $\lambda$, in grand canonical ensemble takes the form:
\begin{equation}\label{mg}
\mathcal{M}=\frac{1}{2} r_{h}+\frac{4}{3} r_{h}^{3} \pi  P+r_{h} \Phi^{2}+\left(32 r_{h} \Phi^{2} \pi  P-\frac{48 \Phi^{4}}{5 r_{h}}+\frac{4 \Phi^{2}}{r_{h}}\right) \lambda
\end{equation}
Furthermore, the black hole temperature and its associated Gibbs free energy up to the first order in $\lambda$ are given by:
\begin{equation}\label{tg}
T=2r_{h} P-\frac{\Phi^{2}}{2 r_{h} \pi}+\frac{1}{4 r_{h} \pi}+\left(\frac{32 \Phi^{4}}{5 r_{h}^{3} \pi}-\frac{48 \Phi^{2} P}{r_{h}}-\frac{2 \Phi^{2}}{r_{h}^{3} \pi}\right) \lambda
\end{equation}
and
\begin{equation}\label{gg}
G=\frac{r_{h}}{4}-\frac{2 r_{h}^{3} \pi  P}{3}-\frac{r_{h} \Phi^{2}}{2}+\left(48 r_{h} \Phi^{2} \pi  P -\frac{56 \Phi^{4}}{5 r_{h}}+\frac{2 \Phi^{2}}{r_{h}}\right) \lambda 
\end{equation}
respectively.

The heat capacity at fixed pressure in terms of the Maxwell potential, is as follows:
\begin{equation}\label{cg}
C = \left[ \left( \frac{\partial \mathcal{M}}{\partial r_h} \right) \left( \frac{\partial T}{\partial r_h} \right)^{-1} \right]_P 
= \frac{
    \frac{1}{2} + 4 r_{h}^2 \pi P + \Phi^2 + 
    \left(
        32 \Phi^2 \pi P + \frac{48 \Phi^4}{5 r_{h}^2} - \frac{4 \Phi^2}{r_{h}^2}
    \right) \lambda
}{
    2P + \frac{\Phi^2}{2 r_{h}^2 \pi} - \frac{1}{4 r_{h}^2 \pi} + 
    \left(
        -\frac{96 \Phi^4}{5 r_{h}^4 \pi} + \frac{48 \Phi^2 P}{r_{h}^2} + \frac{6 \Phi^2}{r_{h}^4 \pi}
    \right) \lambda
}
\end{equation}
Armed with these relations, we are now well-equipped to delve into the thermodynamic properties of this black hole within the framework of the grand canonical ensemble.\par
\begin{figure}[H]
\centering
\subfloat[a]{\includegraphics[width=7cm]{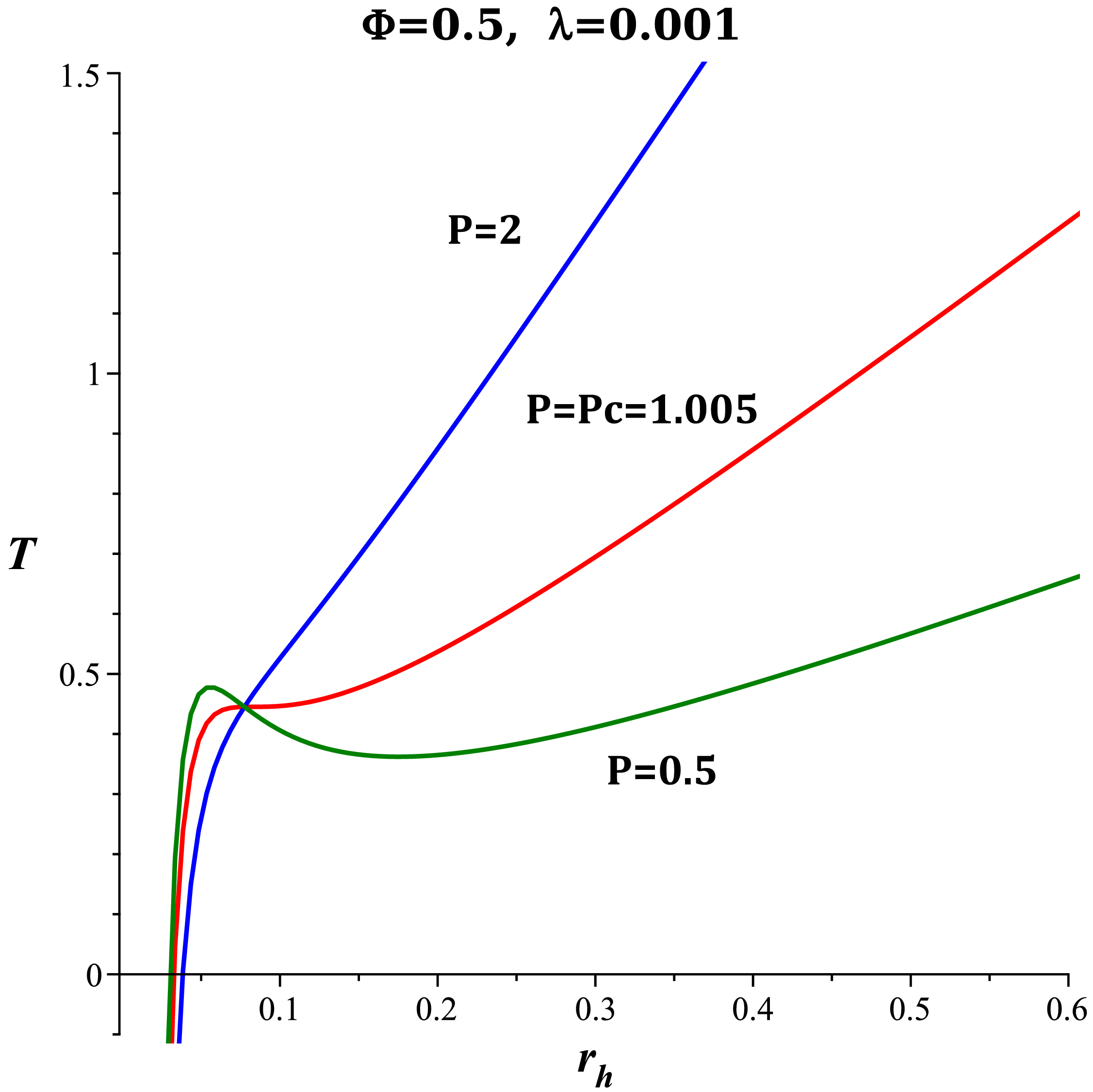}}\quad
\subfloat[b]{\includegraphics[width=7cm]{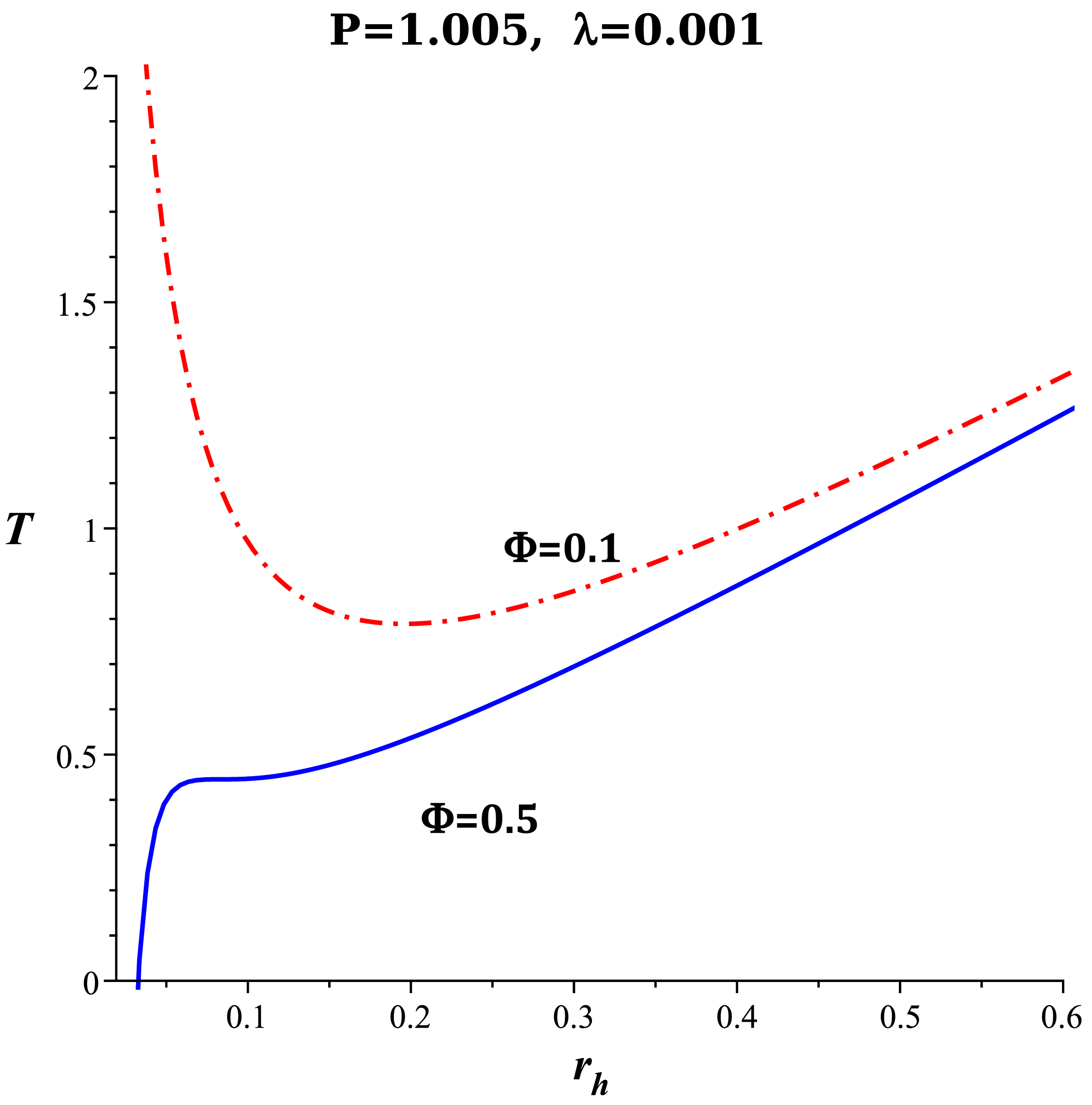}}
\caption{$T$-$r_h$ diagrams of the black hole in the grand canonical ensemble: \textbf{(a)} for varying black hole pressure, and \textbf{(b)} for fixed pressure at the critical value corresponding to the potential $\Phi = 0.5$, while varying $\Phi$. At $\Phi = 0.1$, the temperature behavior approaches that of a Hawking-Page phase transition, exhibiting a critical minimum. \label{fig:9}}
\end{figure}

 
The critical behavior of temperature was investigated but analytical relations can not be obtained. Our analysis demonstrates that the grand canonical ensemble exhibits critical phenomena analogous to van der Waals fluid. Moreover, reducing the potential \( \Phi \) leads to Hawking-Page-like behavior, suggesting a confinement-deconfinement phase transition. This resemblance implies a possible holographic duality between thermodynamic phases in our system and black hole thermodynamics.
The behavior of temperature versus the black hole's horizon for different values of the potential $\Phi$ is illustrated in Fig. \ref{fig:9}. 

 \begin{figure}
\centering
\subfloat[a]{\includegraphics[width=7cm]{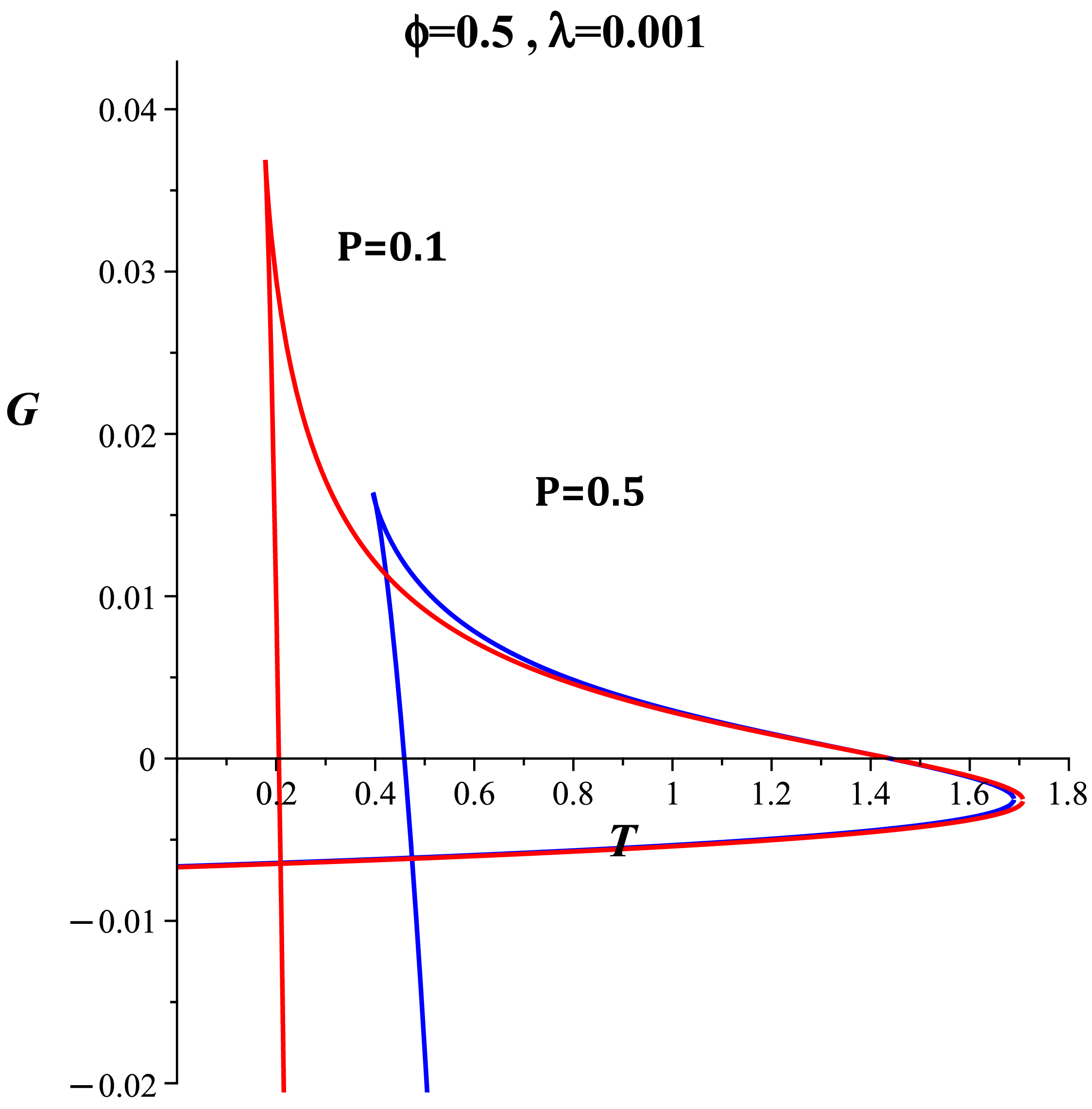}}\quad
\subfloat[b]{\includegraphics[width=7cm]{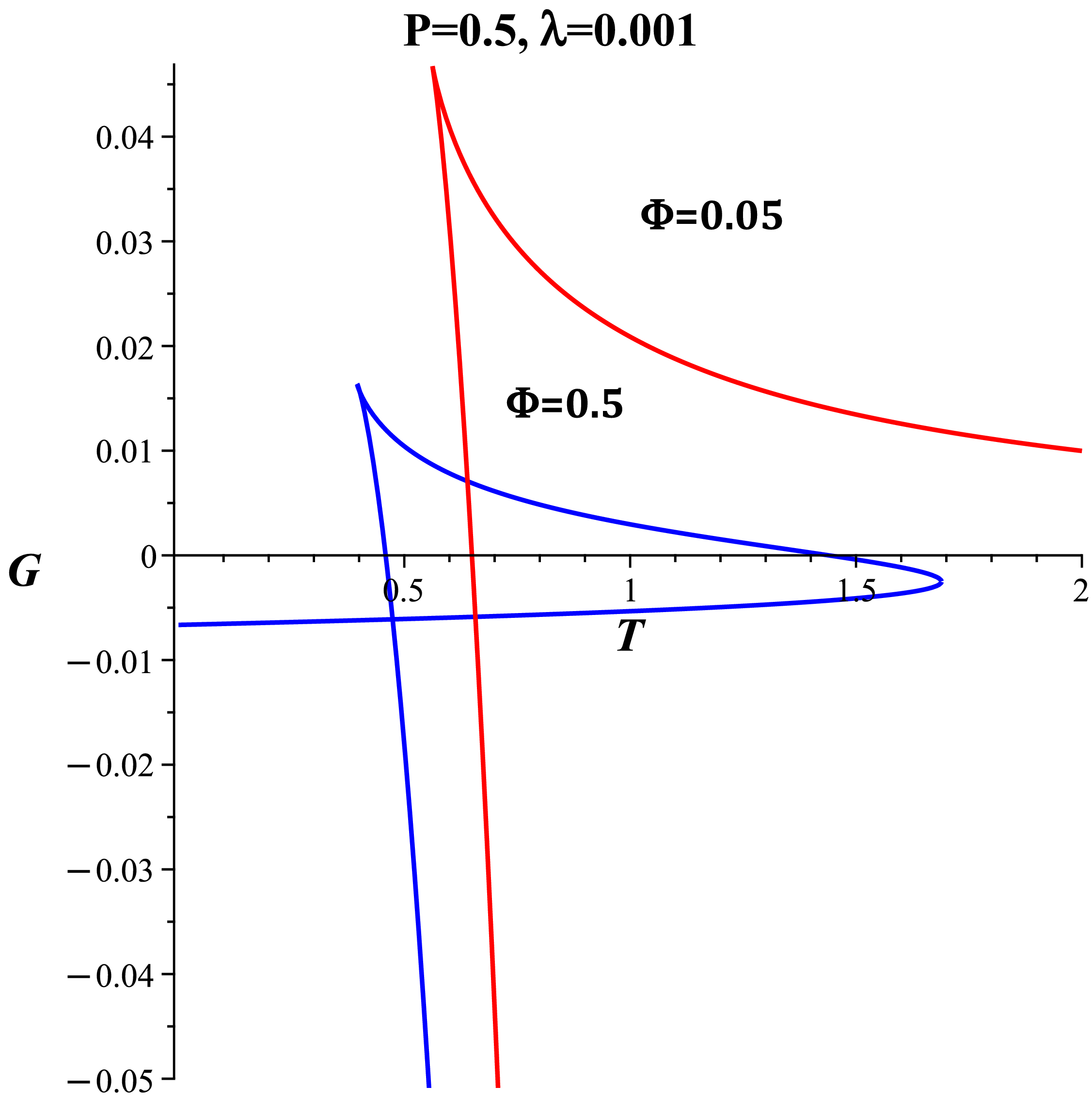}}
\caption{Gibbs free energy ($G$) versus temperature ($T$): \textbf{(a)} Swallow-tail structure, indicative of van der Waals-like phase transitions. \textbf{(b)} Transition between thermal AdS space and black hole states, exhibiting Hawking-Page-like behavior at $\Phi=0.05$. For comparison, the curve corresponding to $\Phi=0.5$ is also plotted.}
    \label{fig:10}
\end{figure}
The study of Gibbs free energy versus temperature reveals that the black hole exhibits van der Waals-like behavior for specific values of the Maxwell potential $\Phi$, corresponding to a small-intermediate-large black hole phase transition. This is evidenced by a swallow-tail shape in the $G$-$T$ diagrams. As the Maxwell potential $\Phi$ decreases, the system displays Hawking-Page-like behavior, signaling a transition between thermal AdS space and the two coexisting states of small and large black holes (see Fig.~\ref{fig:10}). The left panel of Fig.~\ref{fig:10} clearly shows the swallow-tail structure, a hallmark of van der Waals-like systems, while the right panel demonstrates the Hawking-Page-like transition.

\begin{figure}[H]
\centering
\subfloat[a]{\includegraphics[width=7cm]{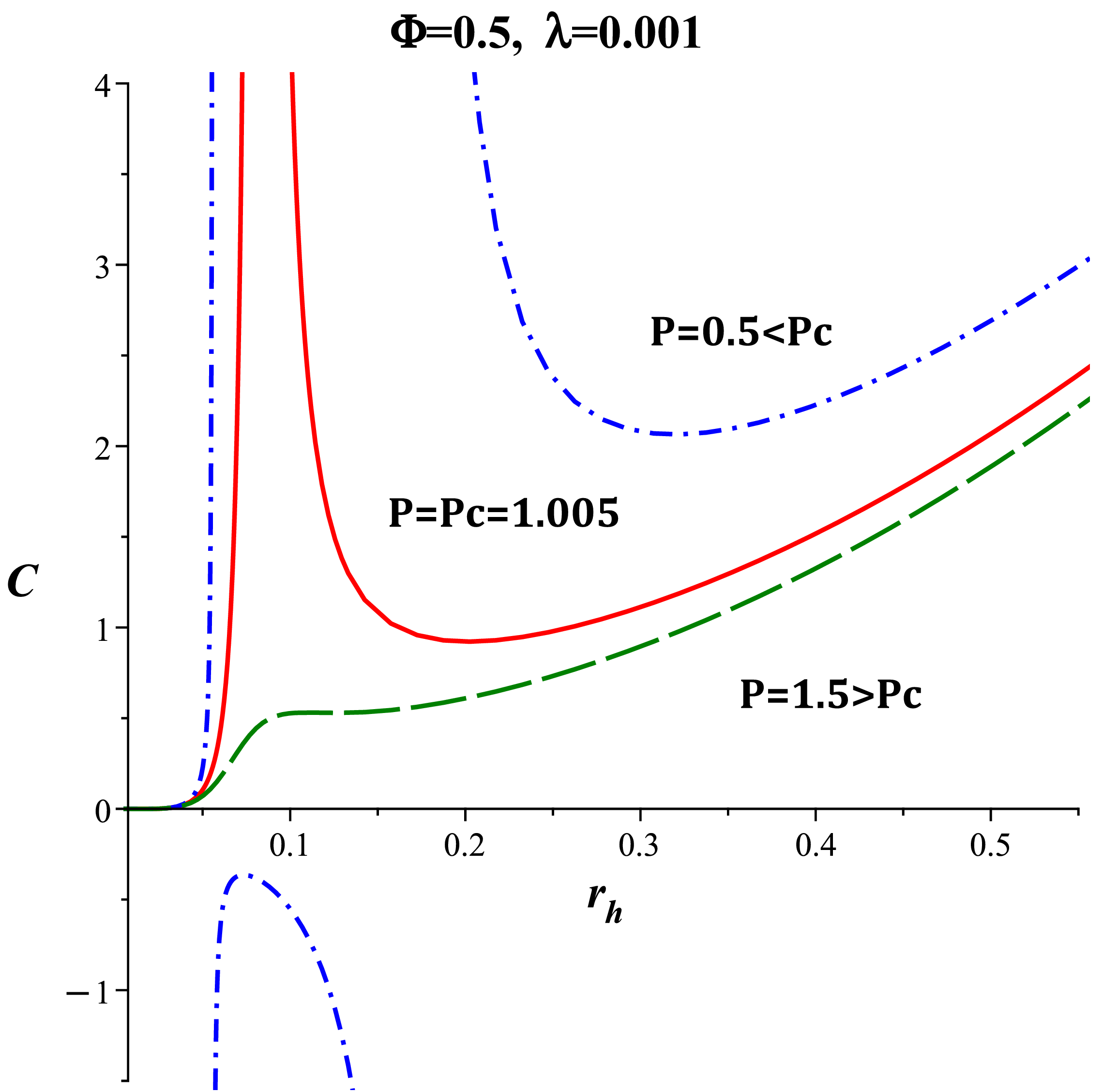}}\quad
\subfloat[b]{\includegraphics[width=7cm]{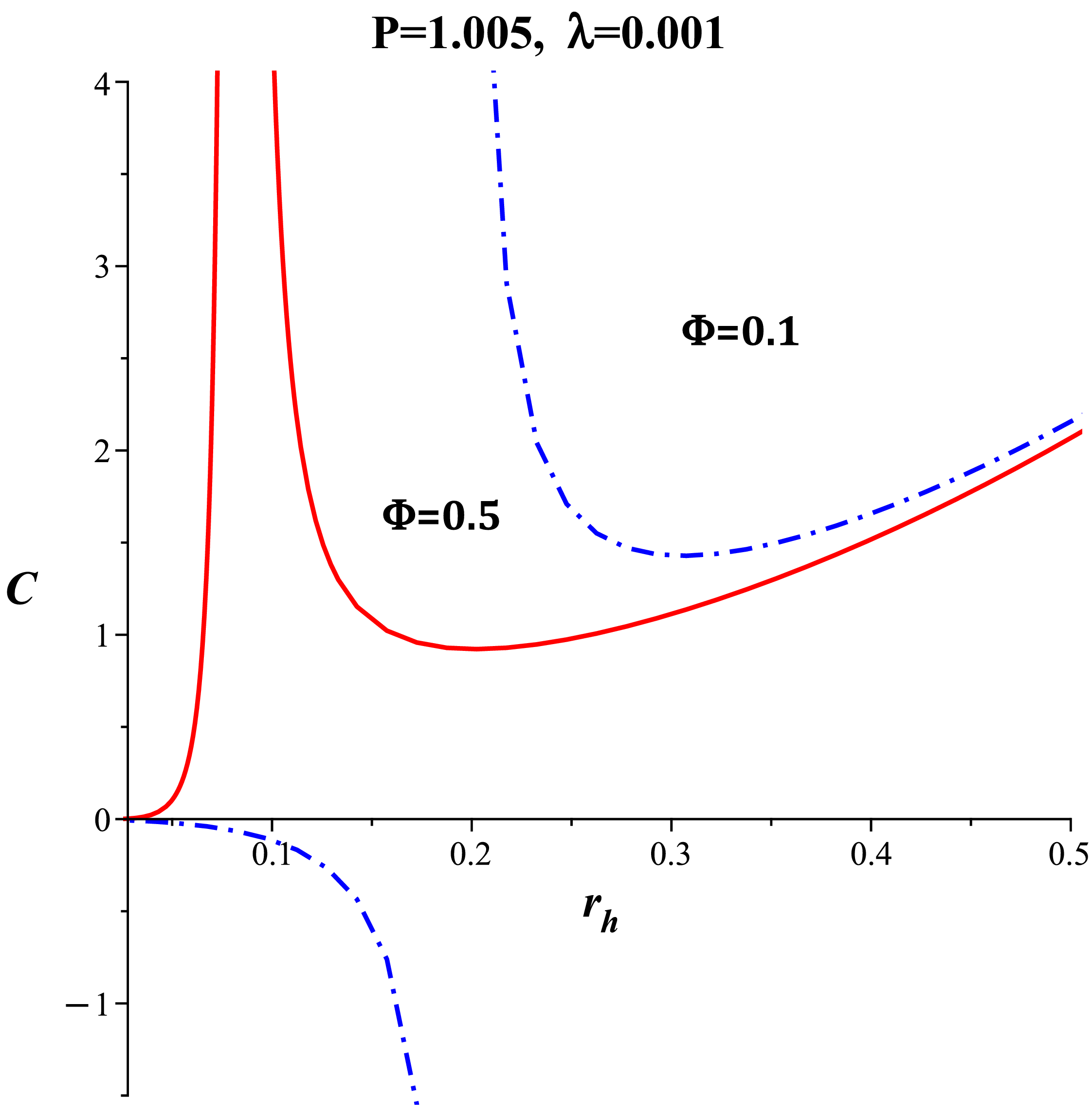}}
\caption{$C$-$r_h$ diagrams of the black hole: \textbf{(a)} For $\Phi=0.5$ and various values of pressure. \textbf{(b)} For fixed pressure and various values of $\Phi$. At the small value $\Phi=0.1$ (compared to $\Phi=0.5$), only two unstable and stable regions exist, exhibiting Hawking-Page-like behavior.} \label{fig:11}
\end{figure}

The investigation of the heat capacity reveals that for some values of the potential $\Phi$, the black hole solution exhibits behavior analogous to a Van der Waals fluid, characterized by distinct phases of stability. Conversely, for small values of $\Phi$, the black hole solution manifests only two stable ($C_P>0$) and unstable ($C_P<0$) regions. These regions correspond to small and large black hole phases, respectively, mirroring the transition behaviors observed in the Hawking-Page-like scenarios. 
In Fig.~\ref{fig:11}, the behavior of the heat capacity is illustrated for various values of $\Phi$. 

Finally, it is interesting to note that black holes can also exhibit thermodynamic behavior in asymptotically flat spacetimes where $ \Lambda = 0 $, under certain conditions. In such situations, the approach of extended phase space does not work, and other methods should be utilized. For instance, flat Lovelock black holes have been studied in Ref.~\cite{RefMann1}, where the zeroth order of the $ N $th-order Lovelock theory of gravity in $ d $-dimensional spacetime is related to the cosmological constant and, consequently, to the pressure of the system. Additionally, Rényi entropy is sometimes utilized for the thermodynamic analysis of black holes in asymptotically flat spacetimes, where the system's pressure is derived from the terms of Rényi entropy \cite{RefProm}. In these cases, phase transitions are observed for black holes.
 \section{Conclusion}
In this study, we have carried out a comprehensive analysis of the thermodynamic behavior, phase structure, and critical phenomena associated with AdS black holes in the presence of a Maxwell field non-minimally coupled to the spacetime curvature. Introducing the coupling constant \( \lambda \), which characterizes the strength of the interaction between the electromagnetic field and the Ricci tensor, we have examined its influence on the black hole solutions and their thermodynamic properties in both the canonical and grand canonical ensembles.

By employing a perturbative approach valid to first order in \( \lambda \), we derived corrected solutions for the metric functions \( f(r) \) and the gauge potential \( h(r) \), incorporating the corresponding integration constants. While in our framework the perturbed event horizon coincides with its unperturbed counterpart at linear order, we emphasize that this behavior is model-dependent. In more general settings, the horizon location may receive first-order or higher-order corrections, a point that deserves further investigation.

We computed the ADM mass (interpreted as enthalpy in the extended phase space), the Hawking temperature, and the Bekenstein--Hawking entropy, revealing that the non-minimal coupling induces explicit \( \lambda \)-dependent modifications. These corrections become especially significant for black holes with small horizon radii, leading to an enhancement in entropy and affecting the thermodynamic stability of the configurations.

In the canonical ensemble, our analysis uncovered a rich Van der Waals-like phase structure, including first-order phase transitions between small, intermediate, and large black holes. The critical behavior, encapsulated in the pressure-temperature relations, Helmholtz free energy profiles, and specific heat, exhibited behavior analogous to classical fluid systems. Furthermore, the divergence points of the heat capacity—typically located near its roots—signal second-order phase transitions, where the system undergoes critical behavior. For small electric charges, the system approaches the well-known Hawking--Page transition, underscoring the interplay between geometry and thermodynamics.

In the grand canonical ensemble, we observed similarly intricate thermodynamic behavior. Both Van der Waals-type and Hawking--Page-like phase transitions emerged, modulated by variations in the electric potential \( \Phi \). These phenomena provide a deeper insight into the phase diagram of non-minimally coupled black holes and reflect the sensitivity of the system to boundary conditions.

From the perspective of the gauge/gravity duality, our results carry profound implications. The Hawking--Page-like transitions at low potential may correspond to confinement--deconfinement transitions in the dual gauge theory, while the Van der Waals-type behavior at higher potentials could be interpreted as analogs of superconducting transitions in holographic models. This dual interpretation not only reinforces the thermodynamic insights obtained from the bulk gravity side but also enriches our understanding of the thermal behavior of strongly coupled field theories.

Overall, our work contributes to the growing body of research at the intersection of modified gravity, black hole thermodynamics, and holography. It highlights how non-minimal couplings can enrich black hole phase structures and deepen the correspondence between gravitational and quantum field theoretic phenomena.

\vspace{1cm}
\noindent {\large {\bf Data Availability} } Data sharing not applicable to this article as no datasets were generated or analysed during the current study.


\end{document}